\documentclass[a4paper,11pt]{article}
\pdfoutput=1
\usepackage{jcappub}

\usepackage{tikz,xcolor,hyperref}
\usepackage{amsmath, amssymb, amsthm, graphicx, epsfig, fancyhdr,epsfig, slashed}
\usepackage[normalem]{ulem}
\usepackage{tikzsymbols}
\usepackage{natbib}
\usepackage{float}
\usepackage{bm}
\usepackage{latexsym}
\usepackage{subfig}
\usepackage{amsfonts}
\usepackage{bigints}
\usepackage{braket}
\usepackage{indent first}
\usepackage{multirow}
\usepackage{blindtext}
\usepackage{titlesec}
\useunder{\uline}{\ul}{}
 \usepackage{environ}
\usepackage{mathtools}
\usepackage{relsize}

\def\o{\omega} 
\def\O{\Omega} 
\def\n{\noindent}

\def\a{\alpha}

\def\k{\kappa}

\def\f{\frac}

\def\lam{\lambda}
\def\mdm{m_\Phi}

\newcommand{\gs}{g_\star}
\newcommand{\gss}{g_{\star s}}
\newcommand{\Trh}{T_\text{rh}}

\newcommand{\tst}{t_\star}
\newcommand{\Tst}{T_\star}

\newcommand{\be}{\begin{equation}}
\newcommand{\ee}{\end{equation}}
\newcommand{\bes}{\begin{equation*}}
\newcommand{\ees}{\end{equation*}}

\newcommand{\beq}{\begin{eqnarray}}
\newcommand{\eeq}{\end{eqnarray}}

\NewEnviron{eqn}{
\begin{align}
\begin{split}
  \BODY
\end{split}
\end{align}
}

\NewEnviron{eqn*}{
\begin{align*}
\begin{split}
  \BODY
\end{split}
\end{align*}
}

\begin{document}
\title{Freeze-in production of scalaron dark matter in $f(R)$ gravity}
\author[a]{Basabendu Barman,}
\author[a]{Ashmita Das,}
\author[a]{Rakesh Kumar SivaKumar,}
\author[a]{and Rudra Pratap Udgata}
\affiliation[a]{\,\,Department of Physics, School of Engineering and Sciences, SRM University AP, Amaravati 522240, India}
\emailAdd{basabendu.b@srmap.edu.in}
\emailAdd{ashmita.d@srmap.edu.in}
\emailAdd{rakesh\_sivakumar@srmap.edu.in}
\emailAdd{rudrapratap\_udgata@srmap.edu.in}
\abstract{We demonstrate that the scalaron, a scalar degree of freedom,  emerging from the $f(R)$ theory of gravity, can account for the observed dark matter (DM) abundance if its mass is around the MeV scale, to ensure its cosmological stability. Focusing on two well-known $f(R)$ gravity models, we systematically show that if scalaron production proceeds via the freeze-in mechanism, the right relic abundance is satisfied over a very narrow window of reheating temperature $10^{14}\lesssim T_{\rm rh}\lesssim 10^{16}$ GeV. We delineate the viable parameter space of the $f(R)$ models consistent with the observed DM abundance, and highlight relevant experimental constraints from searches targeting DM decay signatures.}
\maketitle
\section{Introduction}
\label{sec:intro}
The ordinary matter that makes up stars, planets, and all visible structures accounts for only about 4\% of the total matter-energy budget of the Universe~\cite{Planck:2018vyg}. About 26\% is made up of what is known as the dark matter (DM)~\cite{Bertone:2016nfn, deSwart:2017heh}, whose nature remains one of the most significant open questions in both particle physics and in cosmology. A popular class of candidates are weakly interacting massive particles (WIMPs)~\cite{Steigman:1984ac}, which have been widely studied (see, e.g.,~\cite{Jungman:1995df, Bertone:2004pz, Feng:2010gw}). In the standard WIMP paradigm, the DM is assumed to be in equilibrium with the visible sector, aka the Standard Model (SM), at a temperature much above the DM mass. As the Universe expands and cools, the DM undergoes ``freeze out'', leaving behind a relic abundance that matches the observed DM density, $\Omega_{\rm DM} h^2 \simeq 0.12$~\cite{Planck:2018vyg}. This is an attractive framework since it provides the right DM abundance for a DM-SM interaction strength that roughly matches the weak scale. However, the absence of experimental signals, typically from the direct search experiments, has increasingly challenged this picture~\cite{Roszkowski:2017nbc, Arcadi:2017kky,Arcadi:2024ukq}. Out of several alternative DM production mechanisms, the \emph{feebly interacting massive particles} (FIMPs)~\cite{McDonald:2001vt,Choi:2005vq, Kusenko:2006rh,Petraki:2007gq,Hall:2009bx,Bernal:2017kxu} has gained quite an attention. In the early Universe, FIMPs can be produced via decays or annihilations of the particles in the radiation bath. As the temperature drops below the relevant mass scales, these processes become suppressed, and the FIMP population freezes in, reaching a constant comoving number density. The freeze-in mechanism requires very small interaction rates, either through tiny renormalizable couplings (of size $\sim\mathcal{O} (10^{-10})$), or through non-renormalizable operators suppressed by a high mass scale. The latter, known as ultraviolet (UV) freeze-in~\cite{Elahi:2014fsa}, is especially intriguing because the DM abundance becomes sensitive to the maximum temperature reached by the SM plasma. Under the sudden decay approximation, this maximum is identified with the reheating temperature $\Trh$, marking the beginning of the radiation-dominated era. What is interesting is the fact that the DM-SM interactions can be naturally suppressed by the Planck mass $M_P$, if the effective coupling is composed solely of gravitational interaction at high energies\footnote{One way to realize this is to consider the minimal irreducible interaction that arises from graviton exchange~\cite{Ema:2015dka,Garny:2015sjg,Garny:2017kha,Tang:2016vch,Tang:2017hvq,Ema:2016hlw,Bernal:2018qlk,Ema:2018ucl,Ema:2019yrd,Chianese:2020yjo,Chianese:2020khl,Redi:2020ffc,Mambrini:2021zpp,Barman:2021ugy,Haque:2021mab,Clery:2021bwz,Clery:2022wib,Ahmed:2022tfm} (see also, Ref.~\cite{Kolb:2023ydq} for a recent review), that can successfully reproduce the observed DM abundance.}. 

The $\Lambda$CDM model, often supplemented by an inflationary mechanism driven by a scalar field known as the inflaton, stands as the simplest framework that aligns fairly well with current observational data. Despite its empirical success, $\Lambda$CDM leaves several fundamental questions unresolved—most notably, the origin of the inflaton and the true nature of DM. In addition, it suffers from the well-known cosmological constant problem~\cite{Carroll2001a,Weinberg1989}. While scalar-field models of inflation and dark energy represent modifications to the energy-momentum tensor within Einstein’s equations, an immediate question can be raised that whether the Einstein's general relativity (GR) is sufficient to explain the cosmology of very early Universe. From the fundamental theory such as string/M-theory, it is predicted that GR applies in the region of low curvature and thus represents an approximate theory. Therefore it is legitimate to hypothesise that the early epoch of the Universe was governed by alternative theories of gravity which offer an unified and consistent description of inflation to dark matter to the present day accelerated epoch of the Universe. In the context of cosmological evolution, modified gravity theory should include the terms relevant to the high energy regime of the Universe such as during inflation. As the Universe evolves, the background theory should reduce to the Einstein's GR describing an intermediate phase and lower energy epoch compared to inflation. Subsequently, with the further decrease in the curvature some sub-leading terms in the gravitational action become significant and triggers late time cosmic acceleration. Therefore in the background of the promising modified gravity theories the early to the late time cosmic expansion can be consistently explained solely by the curvature of spacetime and its dynamics. It is worth mentioning that such models should stem from more fundamental theories and are part of the long standing quest for a ``theory of everything" in physics. At this stage we also emphasize that in the present manuscript our prime focus is to explore the DM sector in the background of modified gravity theory. One of the simplest extensions of the modified gravity theory is $f(R)$ theory of gravity, where the Lagrangian density is generalized to an arbitrary function $f(R)$ of the Ricci scalar $R$ (see, e.g., Ref.~\cite{Nojiri2007,Sotiriou2010,DeFelice2010,Nojiri2011} and the references therein). Interestingly, the $f(R)$ gravity can be expressed as the scalar-tensor theory, which includes the Einstein–Hilbert action and an extra scalar field, originating from higher derivatives of $f(R)$. This new scalar degree of freedom, dubbed as ``scalaron,'' could serve as a viable DM candidate as has been investigated in several works~\cite{Capozziello:2006uv,Boehmer:2007kx,Corda:2011aa,Katsuragawa:2016yir,Katsuragawa:2017wge,Yadav:2018llv,Inagaki:2019dqv,Parbin:2020bpp,Shtanov:2021uif,KumarSharma:2022qdf,Shtanov:2022xew,Shtanov:2024nmf}. In these studies, the scalaron is typically considered to be produced either through the misalignment mechanism (oscillating condensate)~\cite{Cembranos:2008gj,Katsuragawa:2017wge,Shtanov:2021uif}, with a mass around the eV scale, or thermally from scattering involving the SM particles in the primordial plasma~\cite{Shtanov:2025nue}, where in the last case it was shown that such a contribution is negligible compared to the energy density of the classical condensate. 

Motivated by these considerations, in this work we adopt $f(R)$ gravity as the underlying theory of gravitation where the scalaron, emerging from the $f(R)$ framework is produced non-thermally from the radiation bath via freeze-in. Upon transforming from the Jordan frame to the Einstein frame, the couplings between the scalaron and the SM fields naturally become suppressed by the Planck scale, ensuring that scalaron production remains out of equilibrium in the early Universe. The conformal transformation also leads to three-point interaction vertices involving one scalaron and two SM particles, which facilitate scalaron decay. By analyzing both tree-level and loop-induced decay channels, we first derive an upper bound on the scalaron mass to ensure its lifetime exceeds (or is comparable) to the age of the Universe, thereby promoting it to a viable DM candidate. On top of that, such decays into SM final states constrain the scalaron mass and lifetime from several experiments that, for instance, search for diffuse gamma or X-rays from the galactic center. Subsequently, we consider two well-motivated models of $f(R)$ gravity as specific examples and explore regions of parameter space consistent with the observed DM abundance. These results, in turn, impose constraints on the corresponding $f(R)$ models. Importantly, modifications to Einstein gravity also alter the Hubble expansion rate through the changes in the field equations, leading to an modified expansion history as discussed, for example, in~\cite{Catena:2004ba,Schelke:2006eg,Donato:2006af,Catena:2009tm,Capozziello:2015ama,Kusakabe:2015yaa}\footnote{Interestingly, such modified Hubble rate can also result in blue-tilted primordial gravitational waves (GW) that may fall within the sensitivity of several futuristic GW detectors~\cite{Bernal:2019lpc,Bernal:2020ywq}.}. We derive the modified Hubble rate in detail and compute the resulting asymptotic DM yield. Our analysis shows that, to achieve the correct DM relic abundance via UV freeze-in, the Universe must have reached temperatures \( \gtrsim 10^{14} \, \text{GeV} \) following inflation, assuming instantaneous reheating and a scalaron with mass $\lesssim\mathcal{O}$ (MeV). This work is entirely devoted in investigating the viability of the scalaron as a DM candidate produced non-thermally via freeze-in, with issues related to the early/late-time acceleration of the Universe and dark energy left unaddressed. 
   
This paper is organized as follows. In Sec.~\ref{sec:setup} we discuss the $f(R)$ theory formalism, followed by Sec.~\ref{sec:model}, where the details of the two $f(R)$ models is provided. In Sec.~\ref{sec:dm}, we elaborate on freeze-in production of scalaron as DM and show the viable parameter space. Finally, we conclude in Sec.~\ref{sec:concl}. The Appendix.~\ref{sec:int} provides scalaron interactions with the SM, which are then utilized to calculate the scalaron decay rates in Appendix.~\ref{sec:DM-decay}. The modification to the Hubble rate in our $f(R)$ models is  detailed in Appendix.~\ref{sec:field}.
\section{Formalism}
\label{sec:setup}
In this section we briefly present the metric formalism for the $f(R)$ gravity theory, where the 4-dimensional action can be written as,
\beq
\mathcal{S}=\f{1}{2\k^2}\bigintsss d^4x\,\sqrt{-g}\,f(R)+\bigintsss d^4x\,\sqrt{-g}\,\mathcal{L}_m\equiv \mathcal{S}_R+\mathcal{S}_m\,,
\label{action_fr_1}
\eeq
where $\k^2=1/M_P^2$ depicts the gravitational coupling strength, $\mathcal{S}_R$ denotes action for modified gravity theory, while $\mathcal{S}_m$ corresponds to the matter action. In this work we follow mostly minus metric signature. The Ricci scalar is defined as $R=g^{\mu\nu}R_{\mu\nu}$, along with the Ricci tensor,
\begin{align}
& R_{\mu\nu}=\partial_\lambda \Gamma^\lambda_{\mu\nu}
- \partial_\mu \Gamma^\lambda_{\lambda\nu}
+ \Gamma^\lambda_{\mu\nu} \Gamma^\rho_{\rho\lambda}
- \Gamma^\lambda_{\nu\rho} \Gamma^\rho_{\mu\lambda}\,,   
\end{align}
where $\Gamma$'s are the usual affine connections defined in terms of the metric tensor $g_{\mu\nu}$ as,
\begin{align}
& \Gamma^{\alpha}_{\beta\gamma} = \frac{1}{2} g^{\alpha\lambda} \left(g_{\gamma\lambda,\beta}
+ g_{\lambda\beta,\gamma}
- g_{\beta\gamma,\lambda}
\right)\,.     
\end{align}
$g$ depicts the determinant of the metric tensor $g_{\mu\nu}$. The Lagrangian $\mathcal{L}_m$ contains all the matter fields, including the Standard Model (SM) fields. 
For the $f(R)$ theory in metric formalism, the action in Eq.~\eqref{action_fr_1}, can be recast 
in the form of Brans–Dicke theory~\cite{Brans:1961sx}\footnote{The $f(R)$ gravity in the metric formalism corresponds to generalized Brans–Dicke (BD) theory with a BD parameter $\omega_{\rm BD}$ = 0~\cite{Chiba:2003ir,PhysRevLett.29.137}.}, where a potential term emerges for the effective scalar degree of freedom. Thus, with the help of an auxiliary field $\chi$, Eq.~\eqref{action_fr_1} can be rewritten as, 
\beq
\mathcal{S}_J=\f{1}{2\k^2}\bigintsss d^4x\,\sqrt{-g}\,\left[f(\chi)+f'(\chi)\left(R-\chi\right)\right]\,, 
\label{action_fr_2}
\eeq
where the subscript ``$J$'' stands for Jordan frame of reference. Varying this action with respect to $\chi$ yields the equation of motion for $\chi$ as,
\beq
f''(\chi)\,\left(R-\chi\right)=0\,.
\label{eom_chi_1}
\eeq 
For $f''(\chi)\neq 0$ one obtains $\chi=R$, which in turn produces the exact form of the action in the Eq.~\eqref{action_fr_1}.  Defining $\phi\equiv\, f'(\chi)$, we write Eq.~\eqref{action_fr_2} as,
\beq
\mathcal{S}_J=\bigintsss d^4x\,\sqrt{-g}\,\left[\f{\phi R}{2 \k^2}- U(\phi)\right],\quad \quad \,{\rm with}\,\,\,\,\, U(\phi)=\f{\chi(\phi)\phi-f(\chi(\phi))}{2 \k^2}\,.
\label{action_fr_3}
\eeq
The action in Eq.~\eqref{action_fr_3} can be further written as,
\beq
\mathcal{S}_J=\bigintsss d^4x\,\sqrt{-g}\,\left[\f{f'(R)\,R}{2 \k^2}-U(\phi)\right],\quad \quad \,{\rm with}\,\,\,\,\, U(\phi)=\f{f'(R)\,R-f(R)}{2 \k^2}\,,
\eeq
where $f'(R)$ is the derivative of $f(R)$ with respect to $R$. Next, we redefine the metric tensor by performing a conformal transformation as below, 
$$g_{\mu\nu}\to \widetilde{g}_{\mu\nu}=\O^2\,g_{\mu\nu},$$ which leads to, 
\beq
R=\O^2\left[\widetilde{R}+6\,\widetilde{\Box}\o - 6 \widetilde{g}^{\mu\nu}\,\partial_{\mu}\o\,\partial_{\nu} \o\right]\,, 
\label{ricci_transform}
\eeq
where $\o\equiv \ln \O$,
$\partial_{\mu}\o\equiv \f{\partial \o}{\partial \widetilde{x}^{\mu}}$, $\widetilde{\Box}\o\equiv \f{1}{\sqrt{-g}}\partial_{\mu}\left[\sqrt{-g}\,\widetilde{g}^{\mu\nu}\,\partial_{\nu}\o\right]$ and $\sqrt{-g}=\O^{-4}\,\sqrt{-\widetilde{g}}$. Due to this transformation, Eq.~\eqref{action_fr_3} becomes,
\beq
\mathcal{S}_J=\bigintsss d^4x\sqrt{-\widetilde{g}}\left[\f{f'(R)\,\O^{-2}}{2 \k^2}\left\{\widetilde{R}+6\, \widetilde{\Box}\o-6 \,\widetilde{g}^{\mu\nu}\partial_{\mu} \o \,\partial_{\nu}\o\right\}\right]\,.
\eeq
The above action reduces to the Einstein frame action under the conditions $\O^2=f'(R)$ and $f'(R)>0$. We further take,  $$\k\, \Phi=\sqrt{\f{3}{2}}\,\ln f'(R),$$ which results in $\o\equiv \ln\O=\k\,\Phi/\sqrt{6}$, with the subsequent action,
\beq
\mathcal{S}_{E}=\bigintsss d^4x \sqrt{-\widetilde{g}}\left[\f{1}{2\k^2}\left\{\widetilde{R} - 6\,\widetilde{g}^{\mu\nu}\,\partial_{\mu}\left(\f{\k \Phi}{\sqrt{6}}\right)\,\partial_{\nu} \left(\f{\k \Phi}{\sqrt{6
}}\right)\right\}-V(\Phi)\right],
\label{action_fr_4}
\eeq
where
\begin{align}\label{eq:vphi-fR}
V(\Phi)=\f{U}{f'(R)^2}=\f{f'(R)\,R-f}{2 \k^2\,\left[f'(R)\right]^2 }\,,  
\end{align}
with $\phi\equiv f'(\chi) \equiv f'(R)=e^{\sqrt{\f{2}{3}}\k \Phi}$. The subscript ``$E$'' refers to the Einstein frame. We finally write down the action in the Einstein frame as,
\beq
\mathcal{S}_{E}=\bigintsss d^4x \sqrt{-\widetilde{g}}\left[\f{\widetilde{R}}{2\k^2} - \f{1}{2}\, \widetilde{g}^{\mu\nu}\,\partial_{\mu}\Phi\,\partial_{\nu} \Phi-V(\Phi)\right]\,,
\label{action_fr_5}
\eeq
along with the matter part, 
\begin{align}
& \mathcal{S}_{E}^{(m)}=\bigintsss d^4x \sqrt{-\widetilde{g}}\,\Bigg[
\widetilde{g}^{\mu\nu} \left(\widetilde{D}_{\mu}\widetilde{H}\right)^{\dagger}\left(\widetilde{D}_{\nu}\widetilde{H}\right) - \frac{1}{4}\widetilde{F}_{\mu\nu} \widetilde{F}^{\mu\nu} + i \bar{\widetilde{f}}\gamma^{\mu}\partial_{\mu} \widetilde{f} + \left(\widetilde{\mathcal{L}}_Y - \widetilde{V}_H\right)
\nonumber \\&
+ \kappa\left\{-\sqrt{\frac{2}{3}}\,\Phi\, \,  \widetilde{g}^{\mu\nu}\,\left(\widetilde{D}_{\mu}\widetilde{H}\right)^{\dagger}\left(\widetilde{D}_{\n u}\widetilde{H}\right)-\sqrt{\f{3}{2}}i\,\left(\Phi\,\bar{\widetilde{f}}\gamma^{\mu}\partial_{\mu} \widetilde{f}-\bar{\widetilde{f}}\gamma^{\mu}\,\partial_{\mu}\Phi\,\widetilde{f} \right)-2\sqrt{\f{2}{3}}\Phi\left(\widetilde{\mathcal{L}}_Y - \widetilde{V}_H\right)\right\}
\Bigg]\,.  
\label{action_ein_1}
\end{align}
From this point onward, we work entirely within the Einstein frame, where the pure Einstein's gravitational action appears and an additional scalar field $\Phi$ emerges due to the higher order curvature effects as introduced by the $f(R)$ theory of gravity. The matter part of the action in Eq. (\ref{action_fr_1}) reduces to Eq.~\eqref{action_ein_1}, where all terms proportional to $\kappa$ give rise to several new interactions between the scalaron ($\Phi$) and all the SM fields in the Einstein frame. This ultimately results in scalaron decay to the SM fields. All the scalaron-SM interaction vertices are reported in Appendix.~\ref{sec:int}.
\begin{figure}
    \centering
    \includegraphics[scale=0.375]{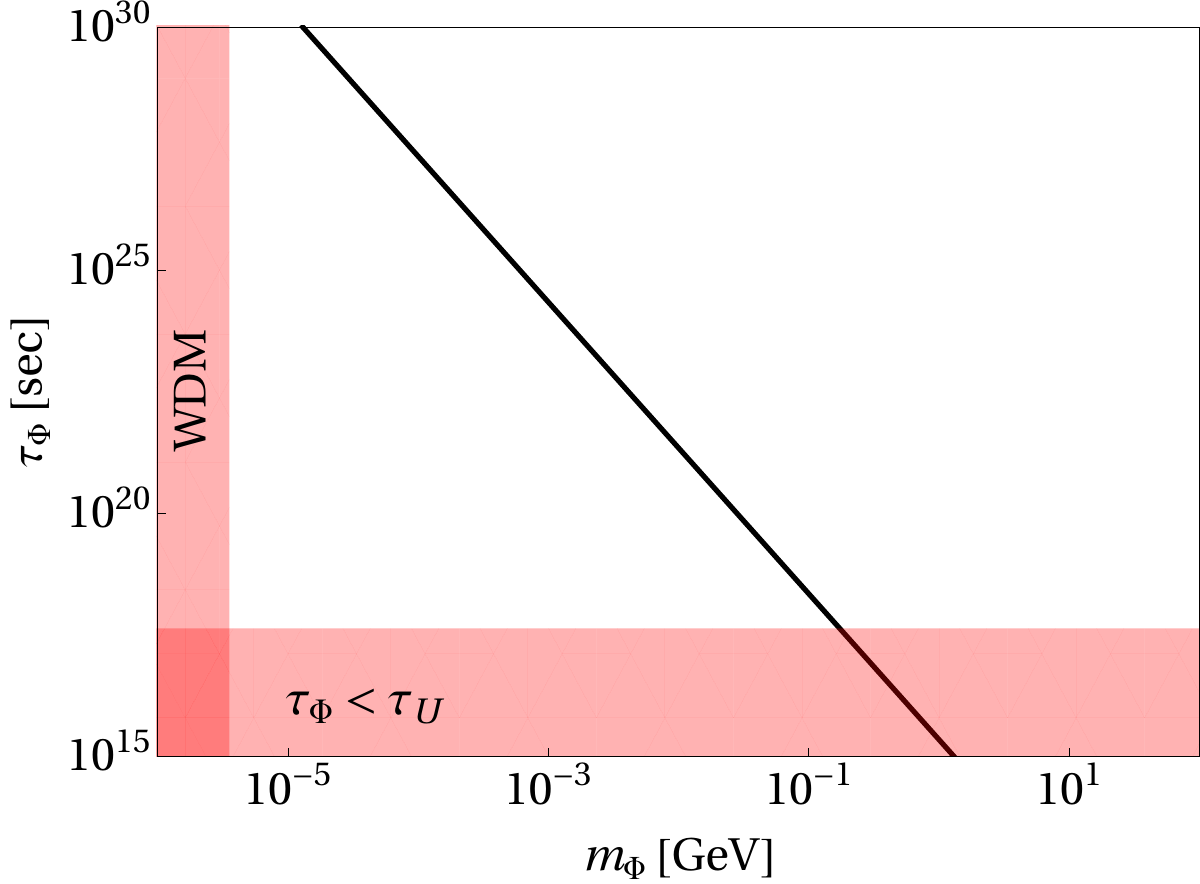}~\includegraphics[scale=0.375]{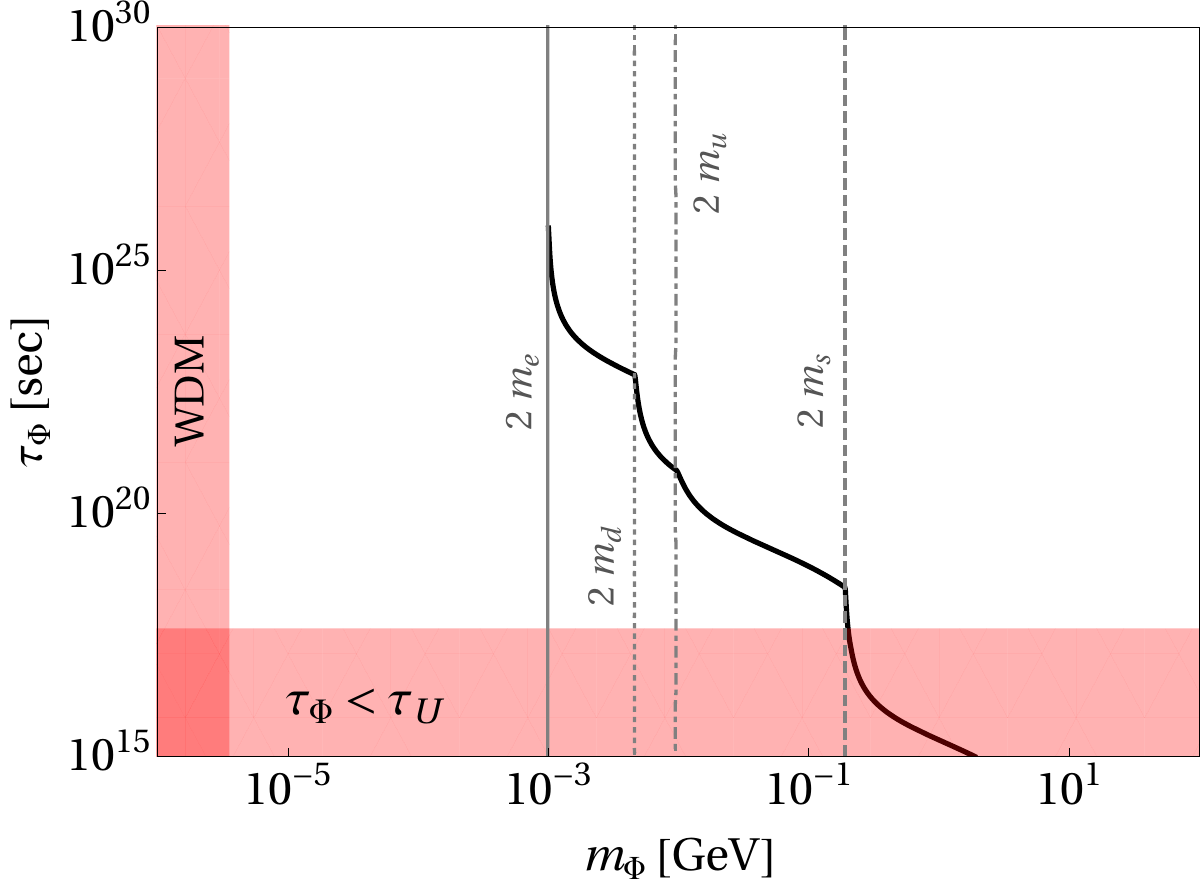}
    \caption{Lifetime of scalaron as a function of its mass before (left panel) and after (right panel) EW symmetry breaking. The horizontal and vertical shaded regions are disallowed from DM lifetime shorter than that of the Universe as well as the warm DM limit, respectively.}
    \label{fig:lifetime1}
\end{figure}

In order to promote the $\Phi$ to a viable DM candidate, two conditions need to be satisfied: (i) the lifetime of $\Phi$ needs to be (at least) comparable to the lifetime of the Universe $\tau_U\simeq 4.3\times 10^{17}$ sec, ensuring the stability of $\Phi$ over cosmological timescale and (ii) $\Phi$ should be able to explain the Planck observed DM abundance~\cite{Planck:2018vyg}. Now, from the action in Eq.~\eqref{action_fr_5}, it is clear that $\Phi$ shall inevitably undergo decay into various SM final states. Hence, we first check the viable scalaron mass range over which its lifetime matches that of the Universe, satisfying condition (i). For that we consider two-body decay rates of the DM into all possible SM states (see Appendix.~\ref{sec:DM-decay} for all relevant decay rates). The bound on DM mass from the requirement of $\tau_\Phi\gtrsim\tau_U$, is shown in Fig.~\ref{fig:lifetime1}. Before the electroweak (EW) symmetry is broken, we consider all the SM fields are absolutely massless. Although in this case the DM can be very light, but still it needs to satisfy $m_\Phi\gtrsim 3.5$ keV, the warm dark matter (WDM) limit~\cite{Irsic:2017ixq,Ballesteros:2020adh}. After EW symmetry breaking, all SM fields become massive. As a consequence, in the right panel, we observe several DM mass thresholds, corresponding to the opening of different decay channels into SM final states. In either cases, we obtain an upper limit of $m_\Phi\lesssim 0.17$ GeV on DM mass, to satisfy the lifetime bound. We emphasize that this bound in independent of the choice of any particular $f(R)$ gravity model, and stems simply from the action itself. Later, we show that this in turn constraints the parameters of a given $f(R)$ model.

We now turn to various observational constraints on the scalaron mass. As pointed out in~\cite{Cembranos:2008gj,Cembranos:2015svp}, a light scalaron induces an additional universal gravitational interaction of Yukawa type. This modifies the total gravitational potential per unit mass to: $V_{\text{grav}}(r) = -\frac{2G}{r} \left( 1 + \frac{1}{3} e^{-\mdm\,r} \right)$. Experimental non-observation of such short-range deviations from Newtonian gravity in tests involving non-relativistic masses imposes a lower bound on the scalaron mass, \(\mdm \geq 2.7 \times 10^{-3}~\text{eV}\), at 95\% C.L~\cite{Kapner:2006si,Adelberger:2006dh,Murata:2014nra,Perivolaropoulos:2019vkb,Shtanov:2024nmf}. Next, observations of both galactic and extragalactic diffuse X-ray and gamma-ray backgrounds impose stringent constraints on decaying DM. In Ref.~\cite{Fischer:2022pse}, lower bounds on the DM lifetime of $\tau_{\rm DM} \gtrsim 10^{21}$ seconds have been established for DM masses ranging from 40~keV to 14~MeV, assuming two-body decays with at least one photon in the final state. In particular, for DM decaying into two photons, a more stringent constraint has been derived in Ref.~\cite{Laha:2020ivk}, yielding $\tau_{\rm DM} \gtrsim 5 \times 10^{26} \left(m_{\rm DM}/\text{MeV} \right)^{-1}~\text{s}$, over the mass range $0.054$--$3.6$~MeV. As mentioned in~\cite{Cembranos:2008gj}, if $\mdm\geq 1.2$ MeV $(\tau_\Phi\leq 1.2\times 10^{24}\,\text{s})$, the scalaron can not constitute the
total local DM since in that case we should observe a bigger excess of the 511 keV line emission of photon coming from the Galactic Center. More broadly, as discussed in~\cite{Essig:2013goa}, for light DM (from a few keV to a few GeV) undergoing radiative decays, the lifetime limits typically span from $10^{24}$ to $10^{28}$ seconds, from HEAO-1~\cite{Gruber:1999yr}, COMPTEL~\cite{Sreekumar:1997yg}, INTEGRAL/SPI~\cite{Bouchet:2008rp,Bouchet:2011fn}, EGRET~\cite{Strong:2004de} and Fermi~\cite{Fermi-LAT:2012edv}. The weaker bounds generally correspond to scenarios involving final state radiation (FSR), while the strongest constraints apply to direct photon emission. Furthermore, energy injection into the intergalactic medium due to DM annihilation or decay can significantly affect the temperature and polarization spectra of the cosmic microwave background (CMB) by altering the post-recombination ionization history. High-energy electrons and photons produced from such processes contribute additional ionization and heating, thereby modifying the CMB signal. This effect has been explored in Refs.~\cite{Diamanti:2013bia,Mambrini:2015sia}, which consider DM decays specifically into electron-positron pairs. For a DM mass of 100 MeV, this places a lower bound on the lifetime of $\tau_{\rm DM} \gtrsim 4 \times 10^{25}$ s. A dedicated study on the constraint on decaying scalaron from diffuse X-/gamma-rays is beyond the scope of the present work, however, we will project the most stringent bounds on the viable parameter space.  
\section{Modified gravity scenarios}
\label{sec:model}
To this end our discussions have been completely independent of the choice of any particular $f(R)$ model, and followed simply the action in Eq.~\eqref{action_fr_5}. We now take up two representative models of $f(R)$. It is clear from Eq.~\eqref{eq:vphi-fR}, the potential of the scalaron depends on the choice of a particular form of $f(R)$, which, in turn, decides the DM mass. Before moving on, it is important to mention that in general, $f(R)$ models must satisfy the following conditions in order to ensure their stability~\cite{Pogosian:2007sw,Miranda:2009rs,DeFelice:2010aj,Guo:2013lka}:
\begin{itemize}
\item [(i)] For the scalaron to remain non-tachyonic, it is required to have $d^2f/dR^2>0$.
\item[(ii)] 
$df/dR>0$ for all finite $R$, to prevent appearance of ghost instabilities. This is simply translated into $e^{\sqrt{2/3}\,\kappa\,\Phi}>0$.
\end{itemize}
In the following subsections we will see, these conditions are going to impose constraints on the choice of parameters for a given $f(R)$ framework.
\subsection{Scenario-A}
\label{sec:modA}
We first consider the following form of $f(R)$,  
\beq
f(R)=R+\a\,R^n\,,
\label{form_fr_1}
\eeq
which includes the Starobinsky’s model~\cite{Starobinsky:1980te} for a specific choice {\it i.e} $(n=2)$. 
Here $\alpha,\,n>0$. Utilizing the fact $f'(R)=e^{\sqrt{\f{2}{3}}\k \Phi}$, we can recast $R$ in terms of the field $\Phi$ as follows,
\begin{align}\label{eq:R1}
& R(\Phi)=\left(n\,\alpha\right)^\frac{1}{1-n}\,\left(e^{\sqrt{\frac{2}{3}}\,\kappa\,\Phi }-1\right)^{\frac{1}{n-1}}\,. \end{align}
Thus one obtains the following form of the scalaron potential from Eq.~\eqref{eq:vphi-fR} as,
\beq
V(\Phi)=M_P^2\,\f{\left(\a\, n\right)^{\f{1}{1-n}}}{2\,n}\,(n-1)\,e^{-\sqrt{8/3}\k\Phi}\left(e^{\sqrt{2/3}\k\Phi}-1\right)^{\f{n}{n-1}}\,.
\label{eq:VmodA}
\eeq
Here $\alpha$ has a mass dimension of $2\,(1-n)$. As discussed before, to avoid tachyonic instability we require
\begin{align}
f''(R)=n\alpha\,(n-1)\,R^{n-2}>0\implies n>1,\,\alpha>0\,.
\end{align}

To understand the behavior of the scalaron potential more clearly, let us first note that its extrema, defined by $V'(\Phi)=0$, occur at those
$R$ that satisfy
\begin{align}\label{eq:1}
    R\,f'(R)-2\,f(R)=0\,,
\end{align}
following Eq.~\eqref{eq:vphi-fR}.
The scalaron mass at such an extremum is given by,
\begin{align}    m_\Phi^{2}\equiv\frac{\partial^2 V}{\partial\Phi^2}\Bigg|_{\Phi=\Phi_{\rm min}}=\frac{1}{3\,f''(R)}\left(1-\frac{R\,f''(R)}{f'(R)}\right)\,.
\end{align}

\begin{figure}[htb!]
    \centering      \includegraphics[scale=0.375]{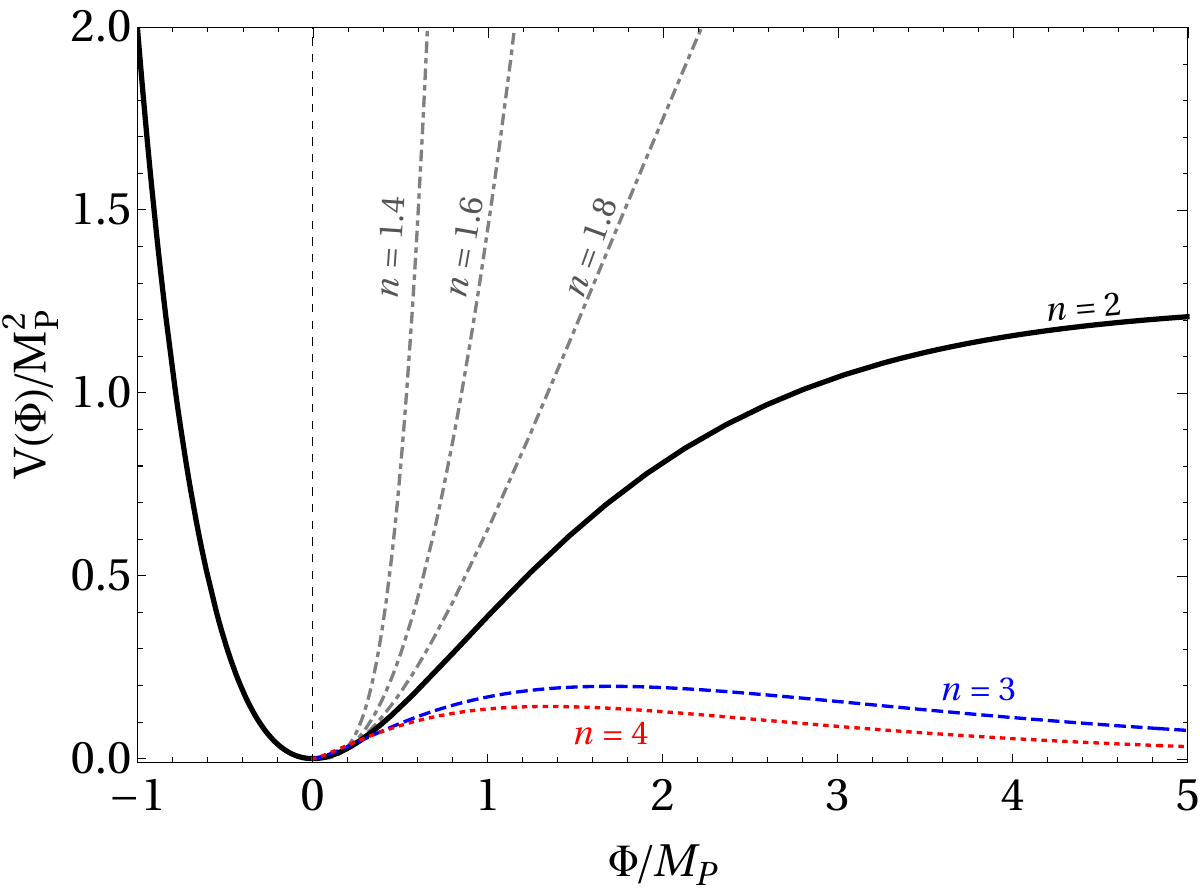}~\includegraphics[scale=0.38]{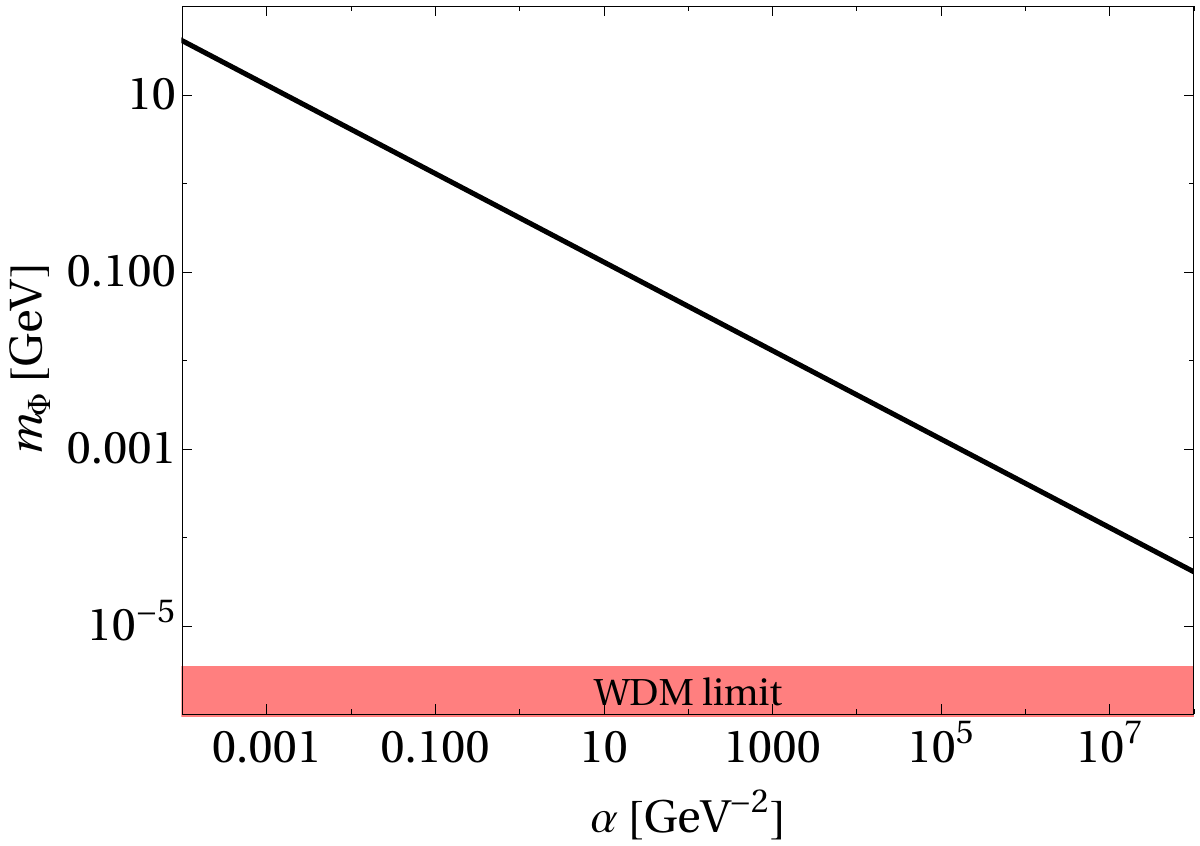}
    \caption{{\it Scenario-A:} The left panel shows the scalaron potential, for different choices of $n$ as shown with different patterns and for a fixed $\alpha=0.1$. Corresponding scalaron mass as a function of $\alpha$, for $n=2$ is shown in the right panel. The red shaded region is disallowed from warm dark matter (WDM) bound.}
    \label{fig:potA}
\end{figure}

Now, for the given power-law model, two cases may arise depending on the values of $n$:
\begin{itemize}
\item For {\it even} $n$, Eq.~\eqref{eq:1} admits two real solutions,
\begin{align}\label{eq:evenn}
    R=0, \qquad
    R^{n-1} = \left[\frac{1}{\alpha\,(n-2)}\right]\,.
\end{align}
Now, for $n=2$ (Starobinsky case) the only possible extremum is $R=0$, which is a regular minimum with corresponding scalaron mass,
\begin{align}\label{eq:scal-mass}
m_\Phi^2 = 1/(6\,\alpha)\,.    
\end{align}
For $n>2$, however, the nonzero root in Eq.~\eqref{eq:evenn} corresponds to a maximum of the potential, while the point $R=0$ becomes a point of inflection rather than a minimum. This can be seen from the left panel of Fig.~\ref{fig:potA}, where we show the scalaron potential as a function of the field values for different choices of $n$, for a fixed $\alpha=0.1$.
\item For {\it odd} $n$, the extremum equation yields three real roots,
\begin{align}\label{eq:oddn}
    R=0, \qquad
    R^{n-1} = \pm\left[\frac{1}{\alpha\,(n-2)}\right]\,.
\end{align}
Here, $R=0$ again corresponds to a point of inflection. The two nonzero extrema form a regular maximum at the positive root and a regular minimum at the negative root, with
\begin{align}
& m_\Phi^2=\frac{2-n}{6n\alpha\,(n-1)}\,R^{2-n}\,.    
\end{align}

The negative root corresponds to an anti-de Sitter space, which is not exactly of our current interest. Note that, $n-1$ is even for odd $n$, and therefore $R^{n-1}>0$ for all real $R$. Consequently, $\Phi\equiv \sqrt{3/2}\,\ln f'(R)>0$ admits solutions only for $\Phi>0$, even though negative curvature is mathematically allowed. This is once again can be clearly seen from Fig.~\ref{fig:potA} for $n=\{3,\,5\}$, shown via blue-dashed and red-dotted curves.
\item Finally, for $1<n<2$, $n-1\in(0,1)$ is fractional. For  negative curvature, $R^{n-1}$ is generically complex. Consequently, $f'(R)$ and $\Phi(R)$ are not real in this regime. Therefore, the scalaron field is defined only for $R\ge 0$, and the corresponding Einstein--frame potential possesses no negative--$\Phi$ branch. These are shown by the gray dot-dashed curves in Fig.~\ref{fig:potA}. 
\end{itemize}
Following the above discussion, we find that for $n=2$ the potential admits a minimum at $\Phi_{\rm min} = 0$, with the corresponding scalaron mass given by $m_\Phi^2= 1/(6\alpha)$. In contrast, for $n>2$, the potential exhibits an extremum located at $\Phi_{\rm ext}
= \sqrt{1/2}\, M_P \,
\ln\!\left[(2n-2)/(n-2)\right]$. In summary, within the class of power-law models $f(R)=R + \alpha R^{n}$, the only phenomenologically viable choice is the Starobinsky case $(n=2)$, for which the scalaron mass is given by Eq.~\eqref{eq:scal-mass}.
\begin{figure}[htb!]
    \centering        \includegraphics[scale=0.375]{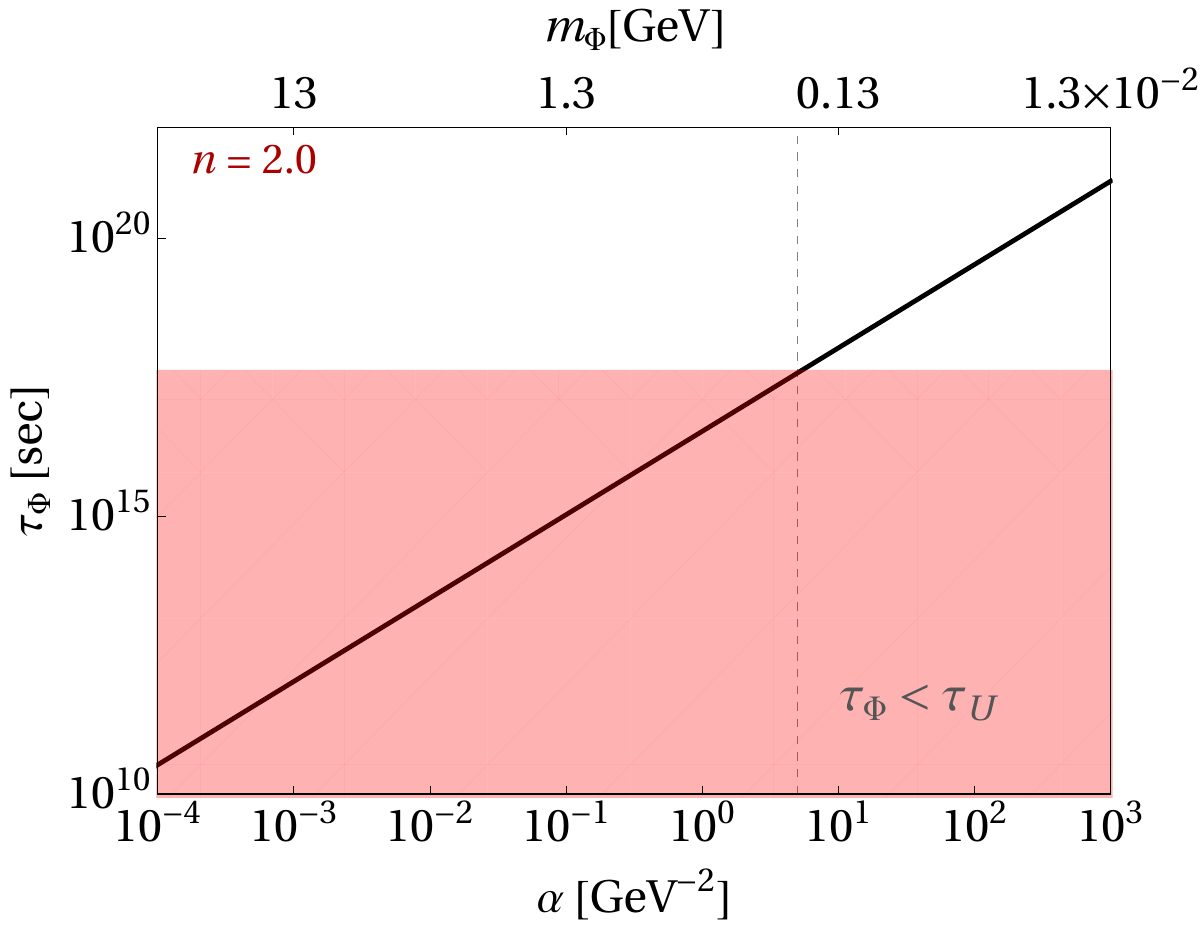}~~\includegraphics[scale=0.375]{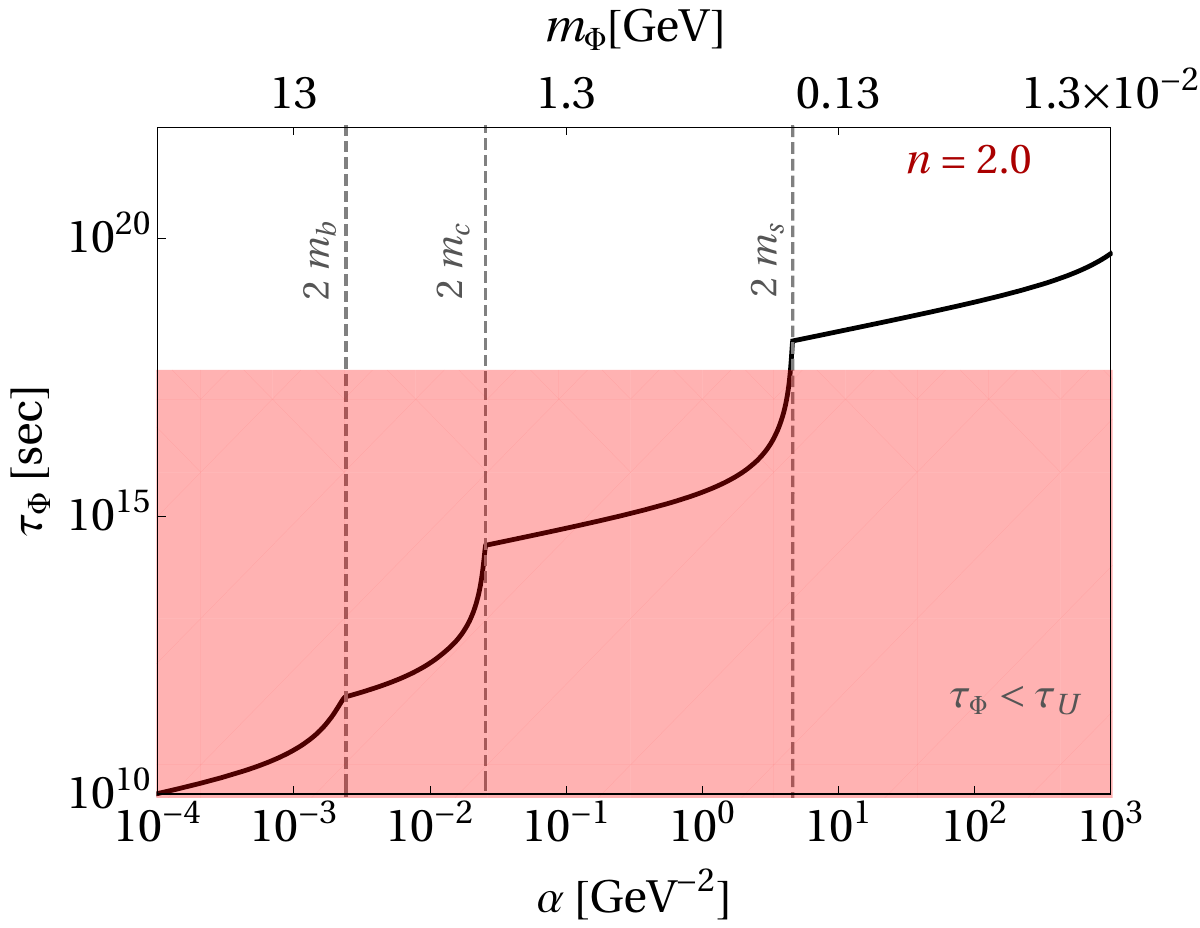}
    \caption{{\it Scenario-A:} Lifetime of scalaron as DM, as a function of $\alpha$. In the left and right panel, respectively, we show scenarios before and after the electroweak symmetry breaking (EWSB).}
    \label{fig:lifetime}
\end{figure}

Since now the scalaron mass is a function of $\alpha$, the requirement of having $\tau_\Phi\propto 1/\mdm^3\propto\alpha^{3/2}\gtrsim\tau_U$, shall provide certain bound on the range of $\alpha$. This is illustrated in Fig.~\ref{fig:lifetime}, for $\alpha\in\left[10^{-3},\,10^3\right]$ and $n=2$. In the left panel we show the scenario where the EW symmetry remains intact, and hence all the SM particles are assumed to be exactly massless. On the other hand, in the right panel, EW broken scenario is considered, and thus we see several mass-thresholds corresponding to $\mdm=2\,m_{\rm SM}$, at which different DM decay channels open up for massive SM final states. Since, $m_\Phi^2\propto 1/\alpha$, hence for larger $\alpha$, the scalaron can have on-shell decay into lighter SM fermions. We find, $\mdm\lesssim 0.17\,\text{GeV}\,\left(0.73/\alpha\right)^3$ such that $\tau_\Phi>\tau_U$.
\subsection{Scenario-B}
\label{sec:modB}
We next consider a logarithmic model of $f(R)$ of the following form
\begin{align}\label{eq:fRB}
& f(R)=R+\gamma\log\left(\frac{R}{\mu ^2}\right)+\lambda\,R^2\,.    
\end{align}
A more generalized version of this model has been considered in~\cite{Nojiri:2003ni}, in order to address late-time acceleration of the Universe. However, as advocated earlier, that is not our concern here. Note that, $\gamma$ has a dimension of mass squared, while for $\lambda$ it is inverse of $\gamma$. Once again, to ensure absence of tachyonic instability,
\begin{align}
& \frac{d^2f}{dR^2}=2\,\lambda-\frac{\gamma}{R^2}>0\implies \gamma<2\,\lambda\,R^2\,.
\end{align}
For $\lambda=0$, the inequality trivially becomes $\gamma<0$. As before, one can express $R$ as a function of $\Phi$ as,
\begin{align}\label{eq:R2}
R(\Phi)=
\begin{dcases}
\frac{1}{4\lambda}\,\left[x-1\pm\sqrt{(x-1)^2-8c}\right]\,, & \lam>0\land \gamma<\frac{1}{8\,\lam}\,\left(1-e^{\sqrt{\frac{2}{3}} \kappa  \Phi }\right)^2\,,
\\[10pt]
\frac{\gamma}{x-1}\,, & \lambda = 0\,,
\end{dcases}
\end{align}
where $x=e^{\sqrt{\frac{2}{3}}\,\kappa\,\Phi}$ and $c\equiv\lambda\gamma$ is a dimensionless quantity. With these, we obtain the scalaron potential as,
\begin{align}\label{eq:VmodB}
& V(\Phi)=\frac{M_P^2}{(z+1)^2}
\begin{dcases}
\frac{1}{16\,\lambda}\,\left[z^2\pm z\,\sqrt{z^2-8c}+4\,c-8c\,\log\frac{z\pm\sqrt{z^2-8c}}{4\lambda\,\mu^2}\right] & \lambda\neq 0\,,
\\[10pt]
\frac{\gamma}{2}\,\left(1+\log \left[\frac{\mu^2\,z}{\gamma }\right]\right)& \lambda = 0\,,
\end{dcases}
\end{align}
where $x=(1+z)$. Only for $+$ sign in the first line of Eq.~\eqref{eq:R2} (along with $\lambda>0,\,\gamma<0$) we obtain a minima in the potential. An analytical expression for the minima of the potential is not possible to compute for $\lambda\neq0$, while for $\lam=0$, one obtains
\begin{align}
\Phi_{\rm min}\simeq\sqrt{\frac{3}{2}}\,M_P\,\log\left[1+\frac{1}{2 \mathcal{W}\left(\frac{0.8\,\mu^2}{\gamma}\right)}\right]\,,
\end{align}
for $\gamma<-\mu^2\,e$, where $\mathcal{W}$ is the Lambert $W$-function. Before proceeding further let us notice that here we have the three independent parameters: $\{\mu,\,\lam,\,\gamma\}$, for $\lam\neq 0$, while for $\lam=0$, this reduces to two. 
\begin{figure}[htb!]
    \centering      \includegraphics[scale=0.375]{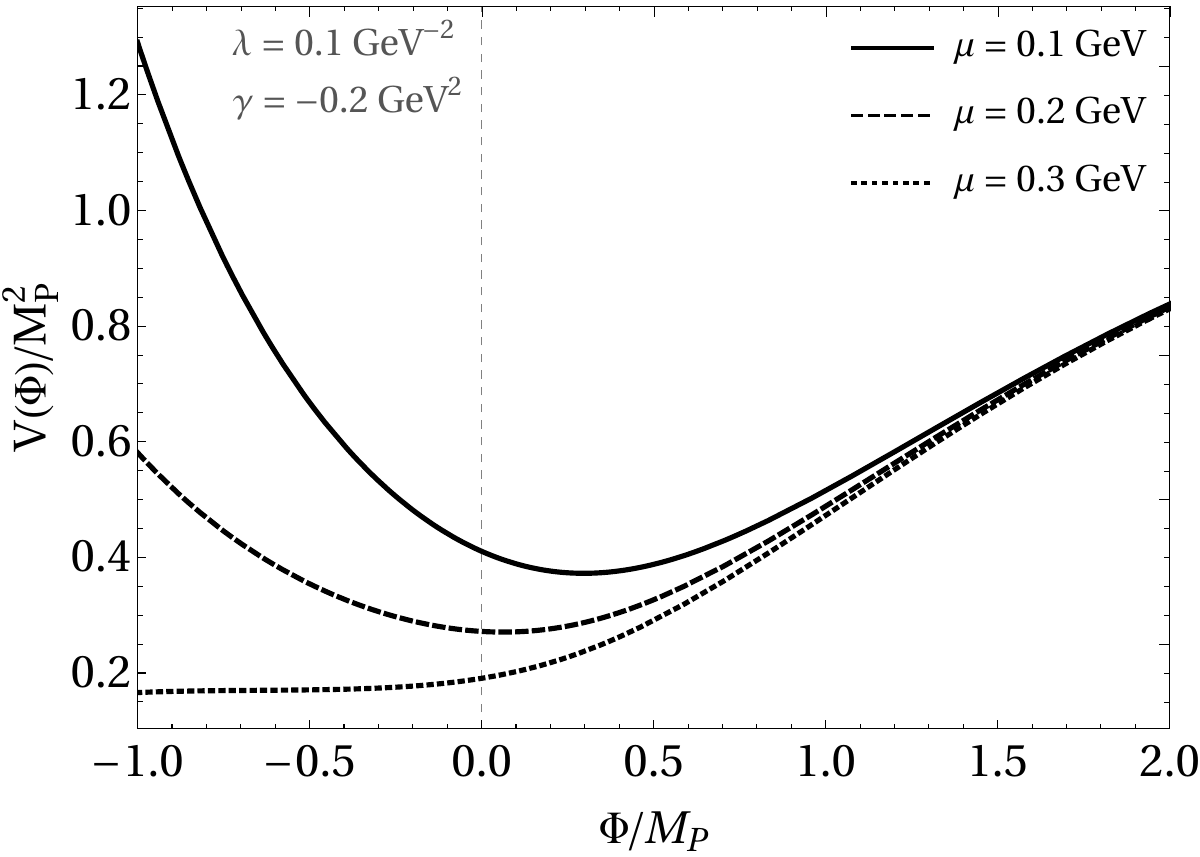}~\includegraphics[scale=0.375]{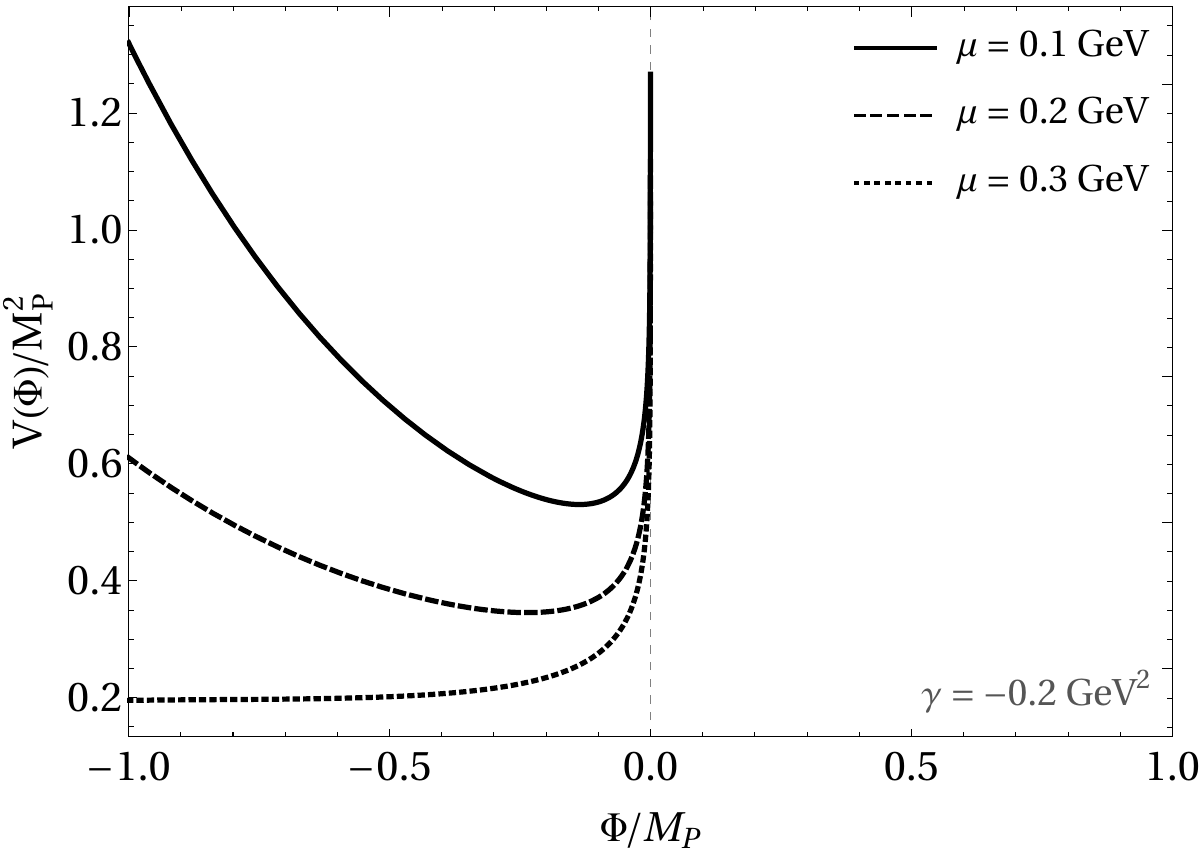}
    \caption{{\it Scenario-B:} The left panel show the scalaron potential for different choices of $\mu$, with a fixed $\lambda=0.1\,\text{GeV}^2$ and $\gamma=-0.2\,\text{GeV}^2$. The right panel shows the same for $\lambda=0$ and $\gamma=-0.2\,\text{GeV}^2$.}
    \label{fig:potB}
\end{figure}
\begin{figure}[htb!]
    \centering    \includegraphics[scale=0.375]{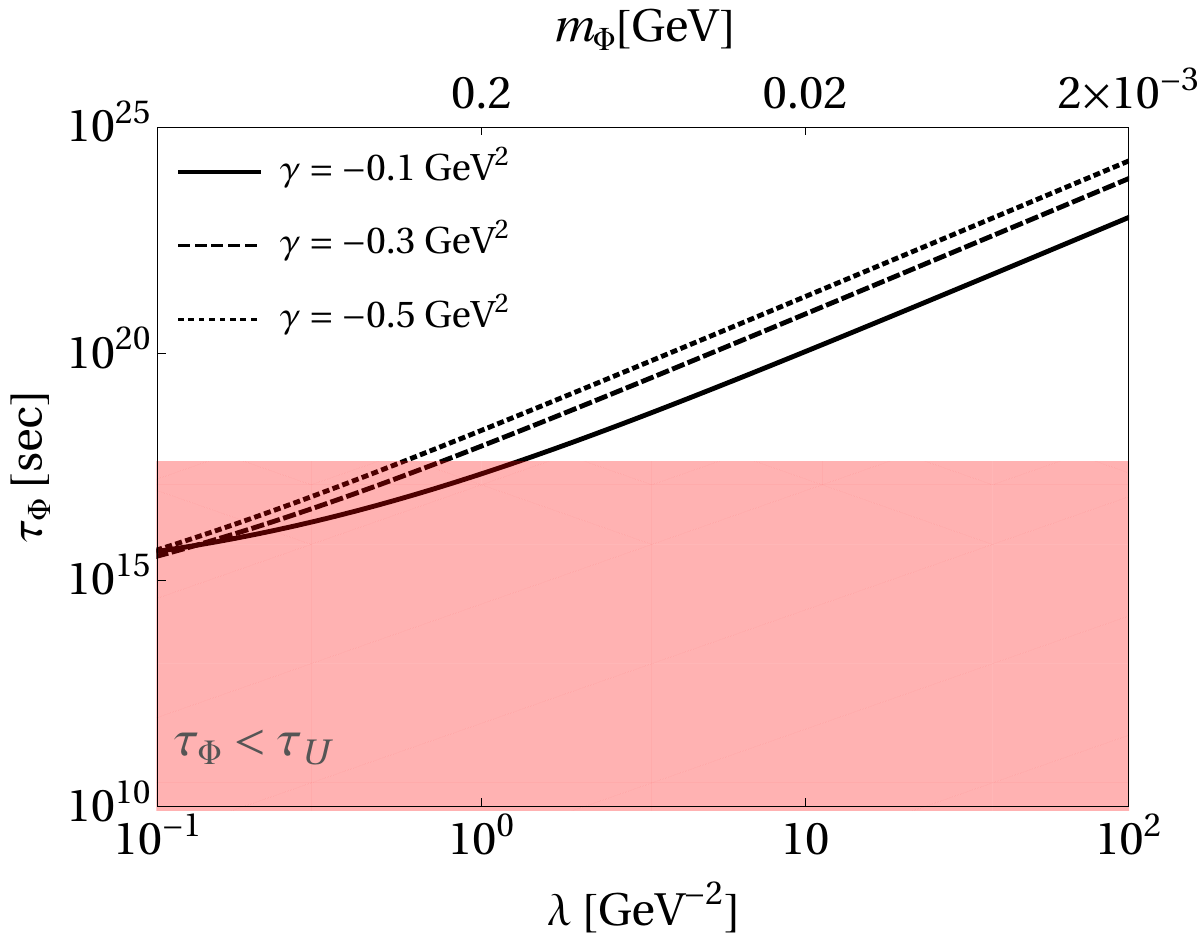}~~\includegraphics[scale=0.375]{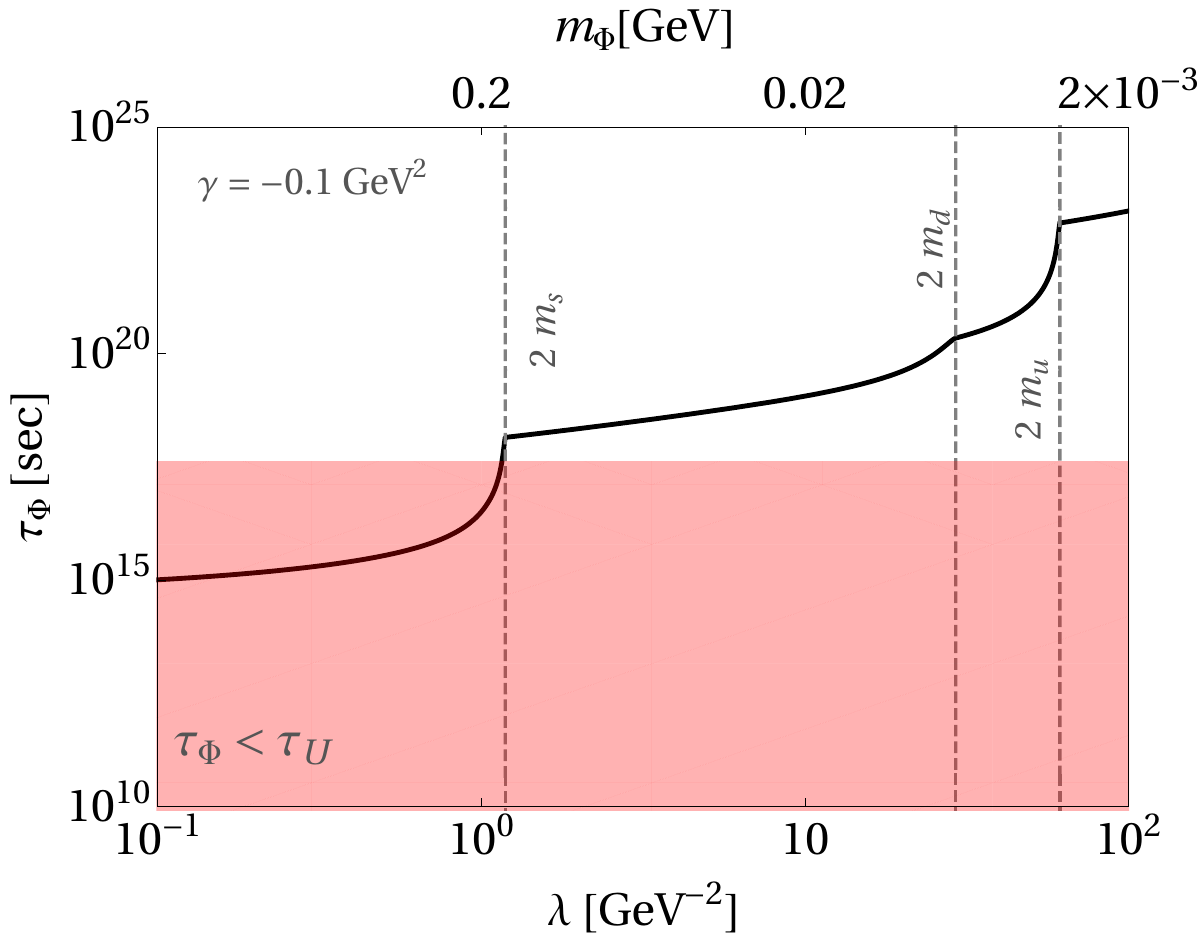}
    \caption{{\it Scenario-B:} DM decay lifetime as a function of $\lambda$ before (left) and after (right) EWSB. In both panels we have fixed $\mu=0.1$ GeV.}
    \label{fig:lifetimeB}
\end{figure}

In the left panel of Fig.~\ref{fig:potB} we show the potential as a function of the field for $\lambda\neq 0$, while the right panel shows the same for $\lambda=0$. For $\lambda>0$ and $\gamma<0$, we see, the minima occurs for $\Phi\geq 0$, depending on the choice of $\mu$, whereas for $\lam=0$, the minima shows up for $\Phi<0$ (right panel). In either cases we obtain a physical mass $(\mdm^2>0)$ for the scalaron. One can find an analytical expression of the scalar mass for $\lam=0$ as, 
\begin{align}
\mdm^2=-\frac{8\gamma}{3}\,\mathcal{W}\left(\frac{0.8\,\mu^2}{\gamma}\right)^2\,\left[\frac{1+\mathcal{W}\left(\frac{0.8\,\mu^2}{\gamma}\right)}{1+2\,\mathcal{W}\left(\frac{0.8\,\mu ^2}{\gamma }\right)}\right]\,, 
\end{align}
which is positive definite for $\gamma<-\mu^2\,e$. The DM lifetime, as a function of $\lam$ is shown in Fig.~\ref{fig:lifetimeB}, considering both before and after EW symmetry breaking scenario. We see, depending on the choice of $\gamma$, it is possible to provide a lower bound on $\lam$, above which the DM is cosmologically stable. This is due to the fact that a larger $\lam$ results in lighter scalaron, making its lifetime even longer, since $\Gamma_\Phi\propto\mdm^3$. For example, with $\gamma=-0.1\,\text{GeV}^2$ and $\mu=0.1$ GeV, one requires $\lambda\gtrsim 1.2\,\text{GeV}^{-2}$ to ensure stability of the DM against decay. This is translated into an upper bound on the DM mass, requiring $\mdm\lesssim 0.08$ GeV. This bound can be relaxed by choosing lower $\gamma$-values, for a fixed $\mu$. Interestingly, for $\lam=0$, we find that the DM lifetime is always below that of the Universe's lifetime. Thus, for {\it Scenario-B}, $\lam=0$ is not a viable option.

At this point we would offer some remarks on the class of $f(R)$ models under consideration. As analyzed in detail in Ref.~\cite{Nojiri:2003ni}, {\it Scenario-B} has the potential to explain the present cosmic acceleration. The inclusion of the $R^2$ term is crucial not only for realizing early-time inflation, but also to satisfy local gravity constraints, such as solar system tests. A similar model was studied in the Palatini formalism in Ref.~\cite{Meng:2003en}, where it was shown to yield an accelerating late-time universe, complying with the conclusion obtained in~\cite{Nojiri2007}. However, unlike {\it Scenario-A}, the presence of a non-zero vacuum energy in this case raises concerns regarding the existence of a Minkowski vacuum solution (existence of a flat spacetime solution). One way to address this issue, as proposed in Ref.~\cite{Amin:2015lnh}, is to modify the argument of the logarithm to the form $\log\left(1 + R/\mu^2\right)$, which ensures $f(R = 0) = 0$. This leads to zero vacuum expectation value for the scalaron, thereby allowing for a flat spacetime solution. In the absence of such a modification, an alternative approach is to introduce a constant term in the $f(R)$ function in such a way that the excessive vacuum energy is canceled and the result matches the observed value of the cosmological constant. Thus it requires some fine tuning through the added constant term to match the current energy density of the universe. As discussed in~\cite{Copeland:2006wr}, this issue 
can be addressed in the background of a more fundamental theory such as string theory or supersymmetry.

\section{Dark matter genesis via freeze-in}
\label{sec:dm}
\begin{figure}[htb!]    \centering\includegraphics[scale=0.14]{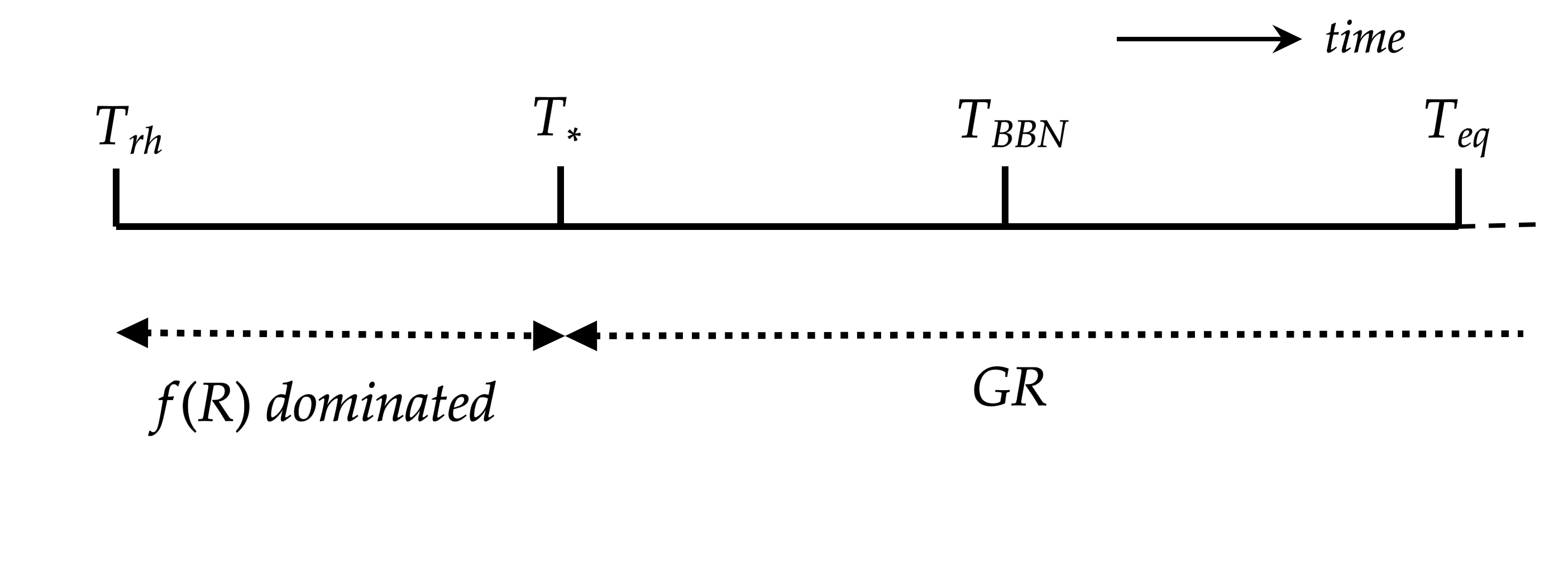}
\caption{A schematic (not to the scale) showing different epochs appearing in the present analysis. Here $\Trh$ is the reheating temperature implying the onset of radiation domination, $\Tst$ is the temperature at which the Universe transits from $f(R)$-gravity to standard GR, $T_{\rm BBN}$ corresponds to the temperature at the onset of BBN and $T_{\rm eq}$ is the epoch of matter-radiation equality. Time flows from left to right, as indicated by the arrowhead at the top.}
    \label{fig:scheme}
\end{figure}

\begin{figure}[htb!]    \centering\includegraphics[scale=0.5]{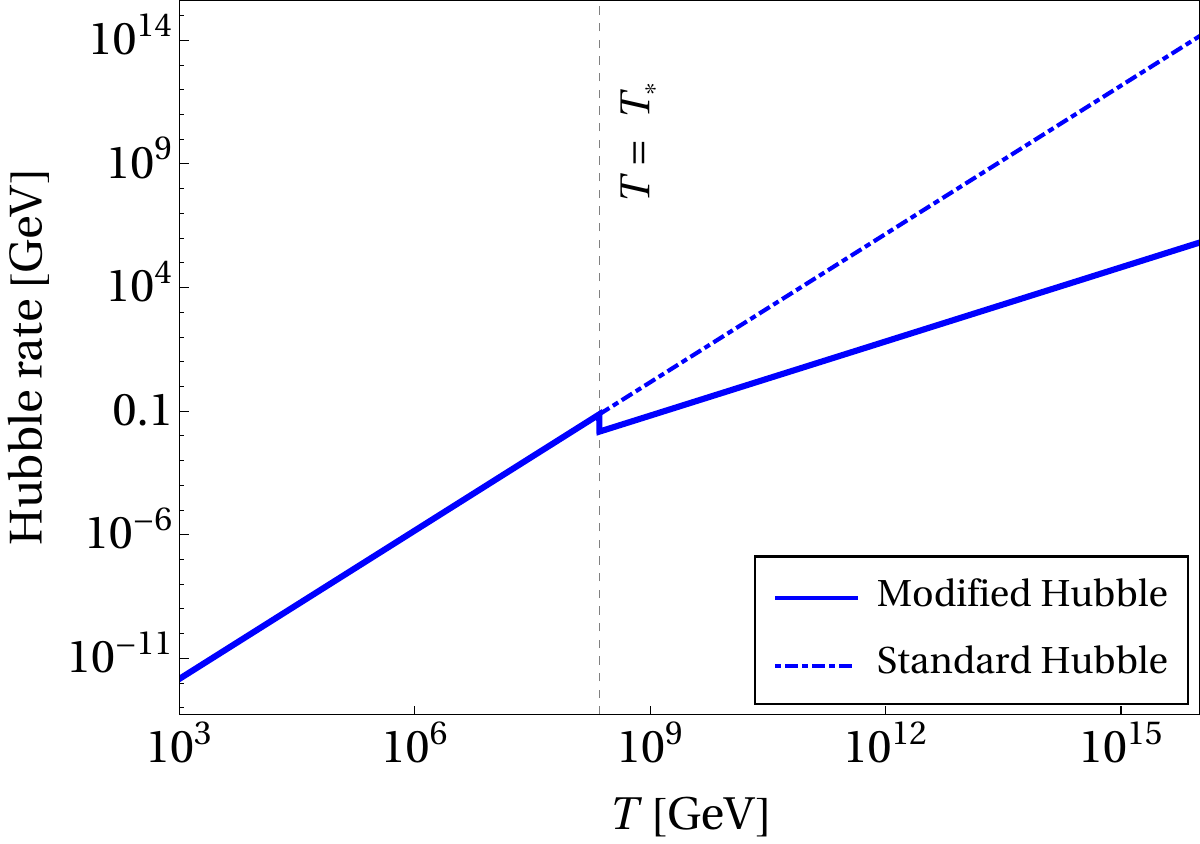}
\caption{Hubble rate as a function of the bath temperature for {\it Scenario-A} with $n=2\,(q=-1)$, where we considered $\alpha=10\,\text{GeV}^{-2}$ and $\beta=0.1$. The solid and dashed curves correspond to modified and standard Hubble during radiation domination, respectively. The vertical dashed straight line corresponds to the transition temperature $T=\Tst$, given by Eq.~\eqref{eq:Tst}. For {\it Scenario-B} (not shown) we obtain similar pattern.}
    \label{fig:hubble}
\end{figure}

Cosmological models that arise from the alternative theories of gravity often predict an early Universe expansion rate  $\mathcal{H}(T)$, that exceeds the Hubble rate $\mathcal{H}_{\rm st}(T)$ of standard cosmology. This deviation can be conveniently captured by introducing a temperature dependent enhancement function $\mathcal{E}(T)$, defined as
\begin{align}
\mathcal{H}(T) = \mathcal{E}(T)\, \mathcal{H}_{\rm st}(T)\,.
\end{align}
We parametrize the enhancement factor as $\mathcal{E}(T) = p\,(T/T_\star)^q$, where $T_\star$ denotes the transition temperature marking the onset of standard cosmological evolution. Although $T_\star$ is a free parameter in principle, it must satisfy the bound $T_\star \gtrsim T_{\rm BBN} \simeq 4~\text{MeV}$ to preserve the successful predictions of Big Bang Nucleosynthesis (BBN). A schematic diagram showing different epochs is shown in Fig.~\ref{fig:scheme}, where for $T\lessgtr\Tst$, we consider a radiation-dominated Universe, until the matter-radiation equality (MRE) $T=T_{\rm eq}$, beyond which matter domination begins. 
The dimensionless constants $p$ and $q$ are determined by the specific form of the underlying $f(R)$ theory. For instance, in {\it Scenario-A}, characterized by $f(R) = R + \alpha\, R^n$, we find $p = 2\beta$ and $q = 2/n - 2 = -1$ as $n=2$, where $\beta$ encodes the relation between the modified scale factor and time (see Appendix~\ref{sec:field} for details). The variation of (modified) Hubble rate with the temperature $T$ is shown in Fig.~\ref{fig:hubble}. For $q=-1$, the modified Hubble rate is reduced compared to the standard Hubble rate during radiation domination for $T>\Tst$, as $\mathcal{H}(T)\propto\mathcal{H}_{\rm st}(T)/T$. Although Fig.~\ref{fig:hubble} corresponds to {\it Scenario-A}, however for {\it Scenario-B} we obtain the exact same nature of the modified Hubble rate, since in that case also we have $q=-1$ [cf. Eq.~\eqref{eq:hub-B}]. 
In a radiation dominated Universe, $\beta$ and $n$ are related via $\beta=n/2$ for {\it Scenario-A} (from Eq.~\eqref{eq:sc-A-trace}), as the trace of the energy-momentum tensor $T=0$. However, here we would consider $T\neq 0$, for which $\beta$ and $n$ can vary independently. This point is further supported by the observation that, in a radiation-dominated Universe with a traceless stress-energy tensor, one obtains $\beta = 1$ for $n=2$ in {\it Scenario-A}. Now, the choice $n = 2$ stems from the specific structure of the underlying $f(R)$ model, which ensures a physical mass for the scalaron and is therefore fixed. However, as noted below Eq.~\eqref{eq:Tst}, maintaining $\Tst > 0$ necessitates $\beta < 1/2$. The resolution lies in introducing an exotic component with an equation of state close to radiation, yet possessing a non-vanishing trace. This radiation-like fluid decouples $\beta$ from $n$ by virtue of its non-zero trace, thereby permitting $\beta < 1/2$ even with $n = 2$. We stress that this exotic fluid is not an additional DM candidate, it only provides a radiation-like background with non-zero trace\footnote{Effect of bulk viscous fluid on particle DM production in a non-standard cosmological scenario has been recently explored in~\cite{Gonzalez:2024dtb,Gonzalez:2024rhs}.}. The scalaron continues to play the role as the sole viable DM component in our framework. Strictly speaking, a non-zero trace challenges the interpretation of the fluid as ``radiation.'' However, we maintain the thermodynamic relation $p=\rho/3$, while allowing for an imperfect fluid in the background\footnote{As shown in Refs.~\cite{Bemfica:2022dnk}, it is possible to realize a model of viscous fluid has no effect on the behavior of radiation, i.e., fluids with an equation of state $w=1/3$.}, that is characterized by an energy-momentum tensor containing, in general, the bulk
viscosity pressure, the energy flux and the anisotropic stress~\cite{Iorio:2014wma}. The system thus mimics radiation from a thermodynamic perspective, while providing richer phenomenology due to modified gravity. Following~\cite{Zimdahl:1997qe,Gron:1990ew,Brevik:2014eya,Capozziello:2015ama}, the bulk viscous pressure can be given by $p_{\rm bulk}=-\epsilon\,\rho/3$, where $\epsilon\equiv \Upsilon(\beta,\,n)/\Delta(\beta,\,n)$, where $\Upsilon(\beta,\,n)$ and $\Delta(\beta,\,n)$ are defined in Appendix.~\ref{sec:app-scenA}. We find, $\epsilon=4\,(1-\beta)/\beta=6$ , for $\beta=0.4$ and $n=2$, implying $p_{\rm bulk}=-6\,\rho/3$.

Due to purely gravitational origin, all DM-SM interactions are naturally suppressed by the Planck scale. This huge suppression prevents the DM to thermalize in the early Universe. Consequently, freeze-out production of DM no more remains a viable option, making freeze-in~\cite{Hall:2009bx,Bernal:2017kxu} a more natural alternative. The key assumption of freeze-in is to consider that the DM had zero or negligible abundance at the earliest epoch, and then the DM yield gradually builds up from the thermal bath with time. In the absence of inverse decay, in the present set-up, DM is produced via 2-to-2 scattering of the bath particles. In order to track the DM number density $n_\Phi$, we solve the  Boltzmann equation (BEQ), that in the present scenario reads~\cite{Hall:2009bx},
\begin{align}\label{eq:beq}
& \frac{dY_\Phi}{dT}=\frac{dY_\Phi}{dT}\Bigg|_{\rm inv. decay}+\frac{dY_\Phi}{dT}\Bigg|_{2\to2}\,,    
\end{align}
where $Y_\Phi\equiv n_\Phi/\mathcal{S}$ represents the DM yield. In the above equation the first term corresponds to contribution from inverse decay, whereas the second term takes into account the 2-to-2 scattering of the bath particles. Now, the contribution from inverse decay is proportional to the DM decay rate $\Gamma_\Phi$, and since $\Gamma_{\Phi}$ needs to be very small in order to ensure the DM stability, one can safely ignore this contribution with respect to the DM yield via scattering. It is possible to write down the BEQ in terms of a dimensionless quantity $x=\mdm/T$ as,
\begin{align}
x\,\mathcal{H}\,\mathcal{S}\,\frac{dY_\Phi}{dx}=\gamma(T)_{2\to2}\,,    
\end{align}
where the entropy density $\mathcal{S}$ and Hubble parameter $\mathcal{H}$ are given by, 
\begin{align}\label{eq:ent-hub}
& \mathcal{S}(T)=\frac{2\,\pi^2}{45}\,\gss(T)\,T^3\,, & 
\mathcal{H}(T)=p\,\left(\frac{\pi}{3}\,\sqrt{\frac{\gs(T)}{10}}\,\frac{T^2}{M_P}\right)\,\left(\frac{T}{\Tst}\right)^q\,,
\end{align}
with $\gss$ and $\gs$ being the effective number of relativistic degrees of freedom. Here,
\begin{align}
& \gamma(T)_{2\to2}=\frac{T}{32\pi^4}\,g_a g_b
\nonumber\\&
\times\int_{\text{max}\left[\left(m_a+m_b\right)^2,\left(m_1+m_2\right)^2\right]}^\infty ds\,\frac{\biggl[\bigl(s-m_a^2-m_b^2\bigr)^2-4m_a^2 m_b^2\biggr]}{\sqrt{s}}\,\sigma\left(s\right)_{a,b\to1,2}\,K_1\left(\frac{\sqrt{s}}{T}\right)\,,
\label{eq:gam-ann}    
\end{align}
is the reaction rate density~\cite{Edsjo:1997bg,Gondolo:1990dk},
where $a,b\,(1,2)$ are the incoming (outgoing) states, $g_{a,b}$ being the degrees of freedom of the initial states, and $K_1[...]$ is the modified Bessel function of the first kind. 
\begin{figure}[htb!]
    \centering        \includegraphics[scale=0.375]{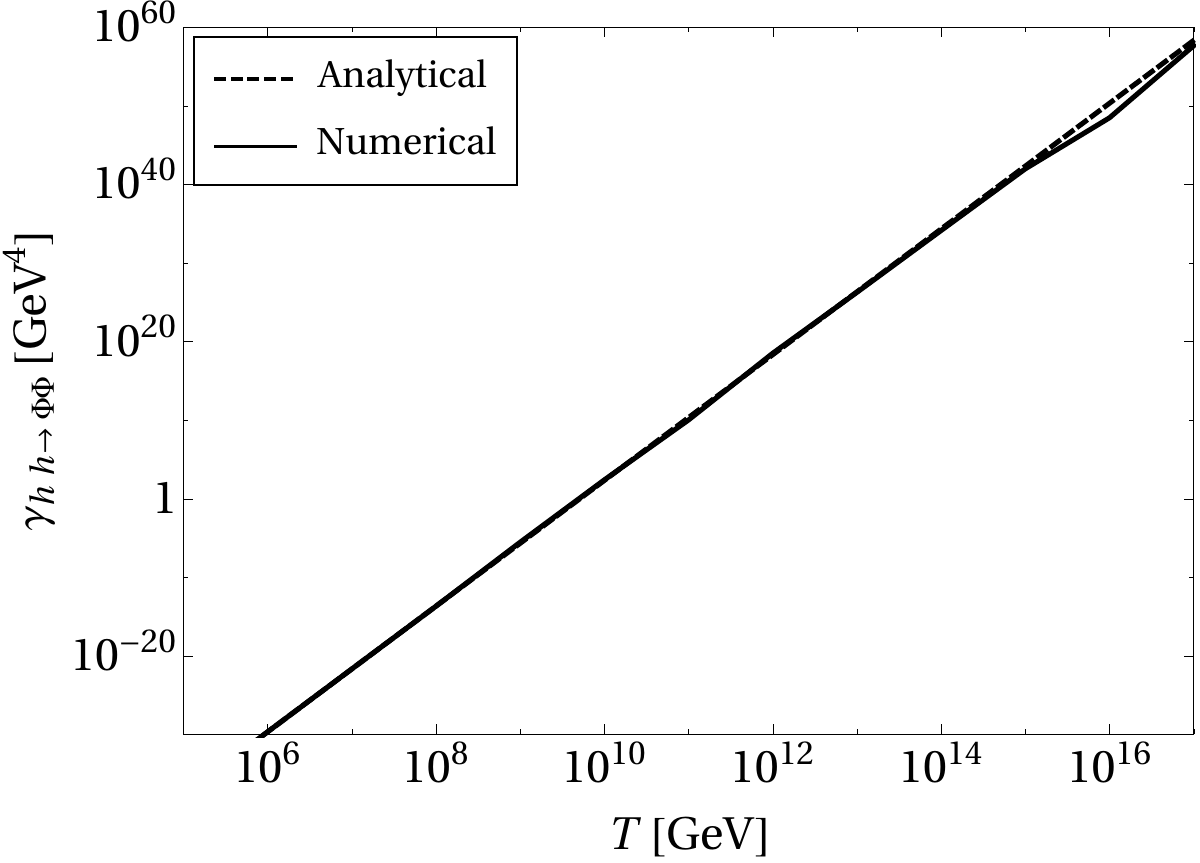}~    \includegraphics[scale=0.375]{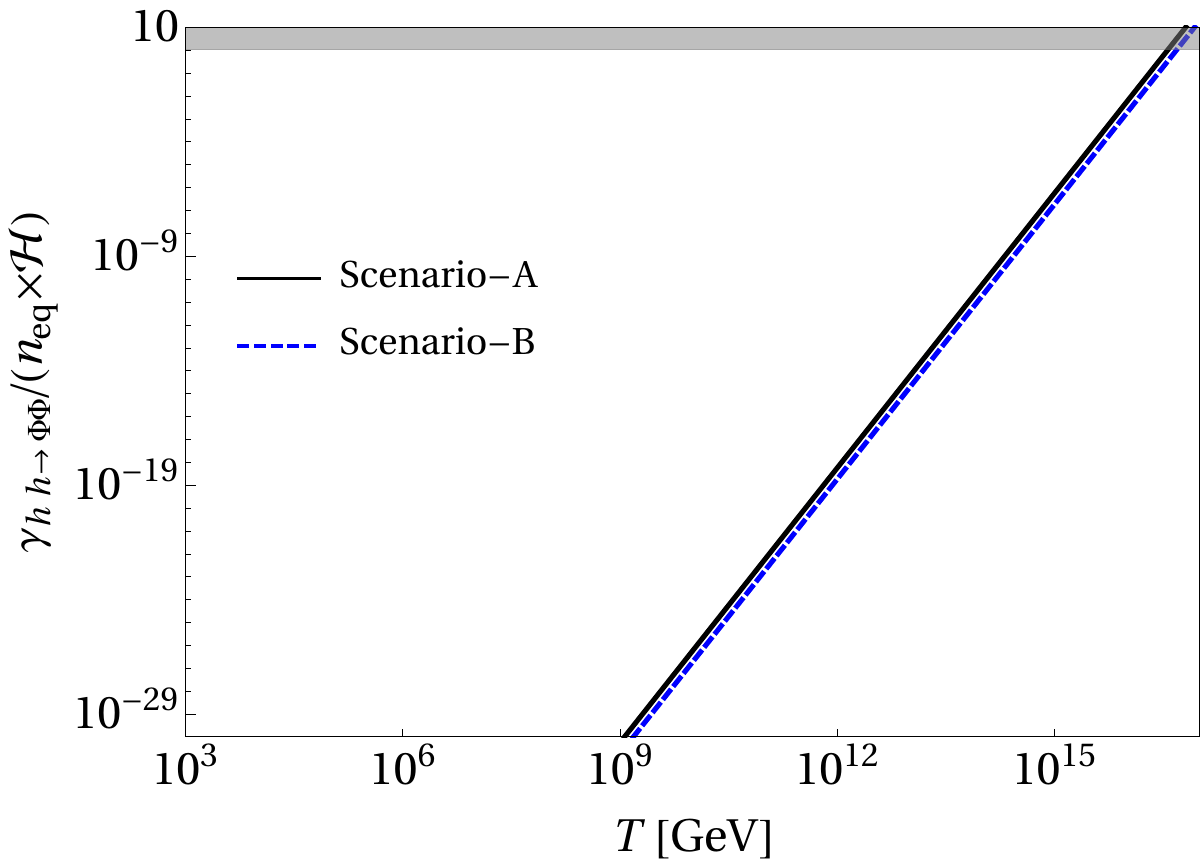}
    \caption{Left: Numerical vs analytical comparison of DM production rate, considering DM of mass 1 MeV. Right: DM production rate vs Hubble rate, as a function of temperature for {\it Scenario-A} (with $\alpha=100\,\text{GeV}^{-2}$) and {\it Scenario-B} (with $\lam=10\,\text{GeV}^{-2},\,\gamma=-0.1\,\text{GeV}^2$ and $\mu=0.1$ GeV), shown via solid black and dashed blue lines, respectively. The top gray shaded region corresponds to DM equilibration with the SM bath, hence forbidden. In both cases $\beta=0.1$ is fixed.}
    \label{fig:gamma}
\end{figure}
\begin{figure}[htb!]
    \centering    \includegraphics[scale=0.52]{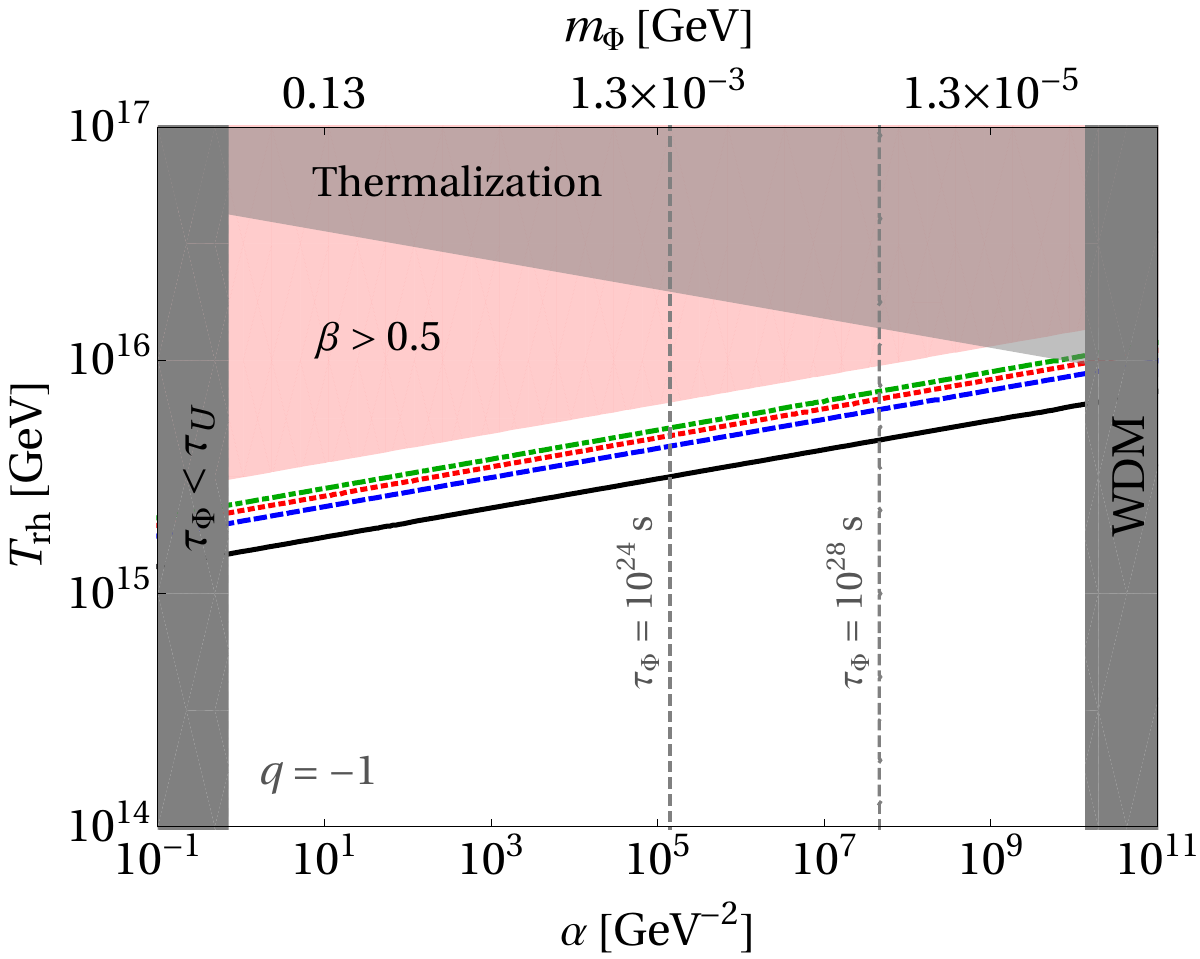}\\[10pt]   \includegraphics[scale=0.52]{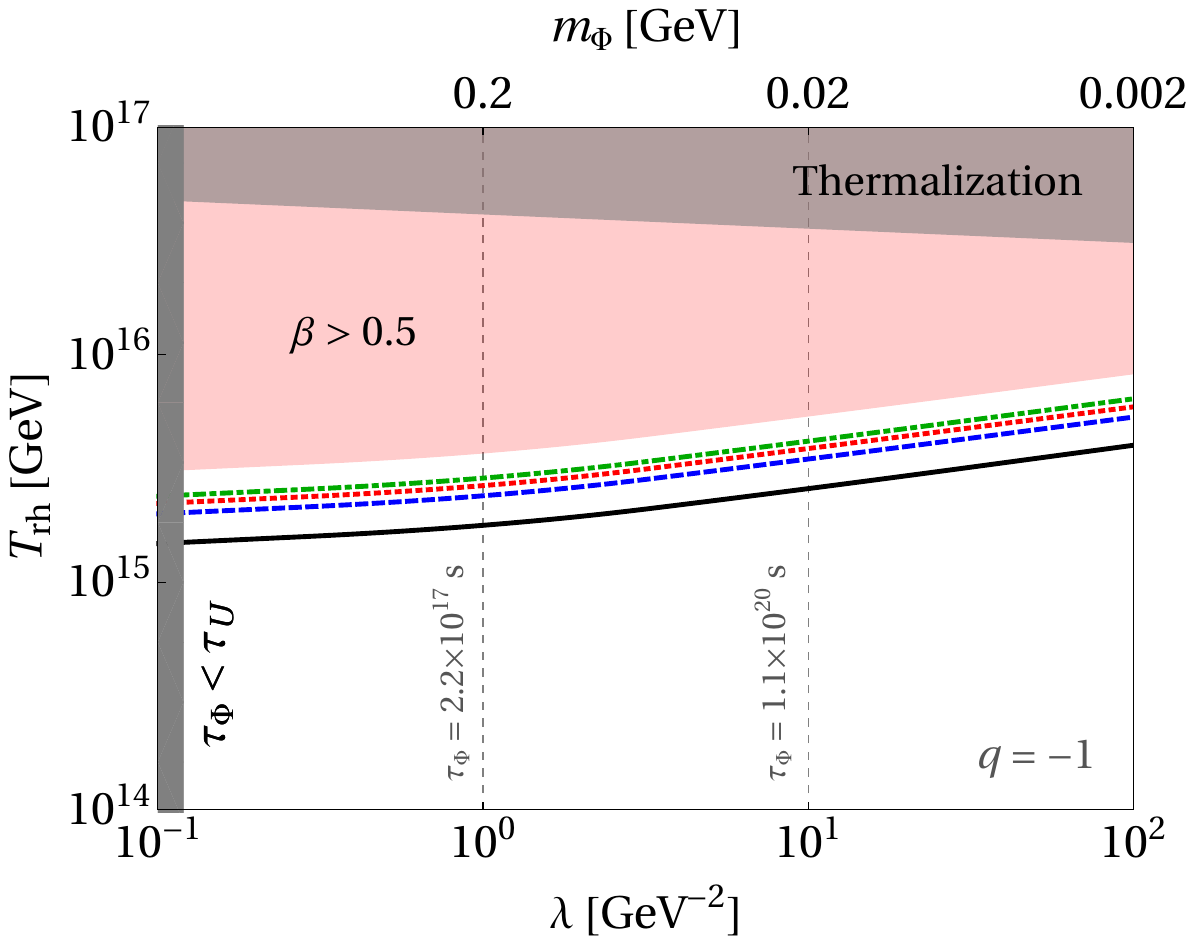}
    \caption{Summary of viable parameter space. Top: {\it Scenario-A}: Contours of right relic abundance in $\Trh-\alpha$ plane corresponding to $\beta=\{0.01,\,0.1,\,0.2,\,0.3\}$ shown by black solid, blue dashed, red dotted and green dot-dashed curves, respectively. Bottom: {\it Scenario-B}: Contours of right relic abundance in $\Trh-\lam$ plane for the same choices of $\beta$ as in the top panel. Here we have fixed $\gamma=-0.1\,\text{GeV}^2$ and $\mu=0.1$ GeV. The vertical dashed lines are bounds on DM lifetime obtained from different observations, as discussed in Sec.~\ref{sec:setup}. Various shaded regions are forbidden from DM lifetime, WDM bound, positivity of $\Tst\,(\beta\ngtr 0.5)$ and DM thermalization (see text for details). }
    \label{fig:paramspace}
\end{figure}

Now, before EWSB, we have the following $t$-channel massless SM Higgs mediated 2-to-2 scattering amplitude,
\begin{align}\label{eq:ampbewsb}
& |\overline{\mathcal{M}}|^2_{hh\to\Phi\Phi}=\frac{s^2}{288\,M_P^4}\,\left[\frac{\left(s-2\mdm^2\right)^2+(s+2\mdm^2)\,\left(2\mdm^2-\cos2\theta\,(s-4\mdm^2)\right)}{\left(s-2 \mdm^2\right)^2-s \cos ^2\theta \,\left(s-4 \mdm^2\right)}\right]^2\,,
\end{align}
where $\theta$ is the scattering angle in the center of mass frame. Since we are considering all the SM states to be exactly massless before EWSB, all other amplitudes vanish. For $T\gg\mdm$, we can ignore the DM mass as well, and further obtain 
\begin{align}
& \sigma(s)\simeq\frac{s}{1152\,\pi\,M_P^4}\,,    
\end{align}
which leads to an approximate analytical expression for the reaction density as,
\begin{align}\label{eq:gamma}
\gamma(T)_{2\to2}\simeq \frac{T^8}{48\,\pi ^5\,M_P^4}\,.   
\end{align}
Note that, the reaction density has a strong dependence on $T$. This is a quintessential feature of UV freeze-in~\cite{Hall:2009bx,Elahi:2014fsa}. As a consequence, in this scenario, bulk of the DM production happens around the maximum temperature of the thermal bath. In the left panel of Fig.~\ref{fig:gamma} we have shown the reaction rate as a function of the bath temperature, considering before EWSB scenario. The reaction rate steadily increases with $T$, as one would expect from Eq.~\eqref{eq:gamma}. Note that, the analytical result (in dashed) matches excellently with the numerical computation (in solid) over a wide range of temperature.  Since freeze-in demands out of equilibrium production of the DM, hence we need to ensure
\begin{align}
& \mathcal{R}\equiv\frac{\gamma_{hh\to\Phi\Phi}}{n_{\rm eq}^{\rm DM}(T)\,\mathcal{H}(T)}<1\,,    
\end{align}
in the early Universe, where $n_{\rm eq}^{\rm DM}(T)=\frac{T}{2\pi^2}\,\mdm^2\,K_1\left(\frac{\mdm}{T}\right)$ is the equilibrium DM number density.
This is shown in the right panel of Fig.~\ref{fig:gamma}, as a function of $T$, for both {\it Scenario-A} and {\it Scenario-B}. Clearly, $\mathcal{R}\ll 1$ over a large temperature range shows the validity of freeze-in production. Here we have considered an equilibrium DM number density, which provides a conservative estimation. The requirement of having $\mathcal{R}(\Trh)<1$, also puts an upper bound on the reheating temperature,
\begin{align}\label{eq:Trh-therm}
\Trh\lesssim 2\times 10^{15}\,\text{Gev}\,\left(\frac{\beta}{0.1}\right)^{1/4}\,\left(\frac{\Tst}{1\,\text{TeV}}\right)^{1/4}\,, 
\end{align}
for $q=-1$. Using Eq.~\eqref{eq:gamma}, we can obtain an analytical solution of Eq.~\eqref{eq:beq} as,
\begin{align}\label{eq:yld}
Y_\Phi (T)\simeq \frac{45}{3392\,\pi^8}\,\sqrt{\frac{5}{53}}
\begin{dcases}
\frac{1}{p\,(3-q)}\,\left(\frac{\Trh}{M_P}\right)^3\,\left(\frac{\Tst}{\Trh}\right)^q\,\left[1-\left(\frac{T}{\Trh}\right)^{3-q}\right]\,,~~q<3\,,
\\[10pt]
\frac{1}{p}\,\left(\frac{\Tst}{M_P}\right)^3\,\log\left(\frac{\Trh}{T}\right) \,,~~q=3\,,
\end{dcases}
\end{align}
which provides the asymptotic DM yield $Y_\Phi(T\ll\Trh)\equiv Y_0$, and we have considered $\gs(T)\simeq \gss(T)\approx 106$. Here, $\Trh$ is the reheating temperature which, in the approximation of a sudden decay of the inflaton, corresponds to the maximal temperature reached by the SM thermal bath. Thus, the DM yield becomes maximum at $T\simeq\Trh$. Finally, to satisfy the observed DM abundance, we demand
\begin{equation} \label{eq:obsyield}
    Y_0\, \mdm = \Omega h^2 \, \frac{1}{s_0}\,\frac{\rho_c}{h^2} \simeq 4.3 \times 10^{-10}\,\text{GeV}\,,
\end{equation}
where $Y_0 \equiv Y_\Phi(T_0)$ is the DM yield today, $T_0\simeq 2.3 \times 10^{-13}\ \text{GeV}
$ being the present day temperature of the Universe. Furthermore, $\rho_c \simeq 1.05 \times 10^{-5}\, h^2$~GeV/cm$^3$ is the critical energy density, $s_0\simeq 2.69 \times 10^3$~cm$^{-3}$ the present entropy density~\cite{ParticleDataGroup:2022pth}, and $\Omega h^2 \simeq 0.12$ the observed abundance of DM relics~\cite{Planck:2018vyg}. By inverting the first line of Eq.~\eqref{eq:yld} we obtain
\begin{align}
\Trh\simeq 4.7\times 10^{14}\,\text{GeV}\,\left(\frac{\Tst}{1\,\text{TeV}}\right)^{1/4}\,\left(\frac{1\,\text{MeV}}{\mdm}\right)^{1/4}\,,    
\end{align}
for $p=2\beta=1,\,q=-1$, in order to saturate the total DM relic. It is worth mentioning that for the same DM mass and $\Tst$, standard radiation domination would require a super-Planckian reheating temperature to produce the right abundance. However, due to modified Hubble rate it is possible to achieve at lower $\Trh$. Since now, because of the $f(R)$ background, the Hubble rate becomes slower (for $T>\Tst$) [cf. Fig.~\ref{fig:hubble}], making the DM production rate more efficient. Note that, the reheating temperature required to produce the right DM abundance is far above the EW transition temperature $(\sim 160\,\text{GeV})$, as a result, most of the contribution to DM yield comes from processes before EWSB. It is worth mentioning here that models of inflation typically suggest an upper bound around $\Trh \lesssim 10^{16}\,\mathrm{GeV}$ (see, for example, Ref.~\cite{Linde:1990flp}). In particular, a large $\Trh$ can potentially lead to the overproduction of long-lived exotic relics, which may overclose the Universe. A prominent example of such an issue is the cosmological gravitino problem in supergravity theories~\cite{Moroi:1993mb}, that leads to an upper bound\footnote{It has been demonstrated in~\cite{Cyburt:2002uv,Kawasaki:2004yh} that for gravitino masses of order 100~GeV, the constraint on the reheating temperature becomes even more stringent, requiring $\Trh \lesssim (10^6\text{--}10^7)$~GeV.} of $\Trh \lesssim 10^{10}$~GeV. To keep the BBN predictions unharmed, we require $\Trh\gtrsim 4$ MeV~\cite{Sarkar:1995dd, Kawasaki:2000en,Hannestad:2004px, DeBernardis:2008zz, deSalas:2015glj,Hasegawa:2019jsa}, which provides a lower bound on $\Trh$.

The relic density allowed parameter space for {\it Scenario-A} is illustrated in the top panel of Fig.~\ref{fig:paramspace}. Note that, here we have following set of independent variables, over which we will scan to obtain the viable parameter space: $\{\alpha,\,\beta,\,\Trh\}$. First of all, there is a lower bound on $\alpha$ coming from the requirement of $\tau_\Phi>\tau_U$, while the upper bound is arising from WDM limit of $m_\Phi\gtrsim 3.5$ keV. Putting these together, we obtain,  $0.67\lesssim\alpha\lesssim 1.3\times 10^{10}$. From the first line in Eq.~\eqref{eq:yld}, we find the asymptotic DM yield as
\begin{align}\label{eq:yld-A}
& Y_0=\frac{45}{27136\,\pi^8\,\beta}\,\sqrt{\frac{5}{53}}\,\left(\frac{\Trh^4}{M_P^3 \Tst}\right)\,,  
\end{align}
where we have used $q=-1$, corresponding to $n=2$. The above expression clearly shows that the DM yield is highly sensitive to the reheating temperature. Moreover, it is evident that $\beta \not< 0$. The transition temperature $\Tst$ is determined by [cf. Eq.~\eqref{eq:Tst}],
\begin{align}\label{eq:Tstn2}
& \Tst\big|_{n=2}\simeq 2\times 10^8\,\text{GeV}\,\sqrt{\frac{0.1}{\beta}}\,\left(\frac{10\,\text{GeV}^{-2}}{\alpha}\right)^{1/4}\,,
\end{align}
implying $\Tst$ decreases with increase in $\alpha$ [cf. left panel of Fig.~\ref{fig:Tst}]. Along with Eq.~\eqref{eq:yld-A}, we find, the DM yield becomes $Y_0\propto \left(\Trh/\alpha\right)^{1/4}$. Consequently, higher $\Trh$ is required to satisfy the observed DM abundance with increase in $\alpha$, for a given $\beta$. However, $\Trh$ cannot be arbitrarily large, as doing so would violate Eq.~\eqref{eq:Trh-therm}, leading to the thermalization of DM. This constraint is shown by the upper gray shaded region for $\beta = 0.01$, which yields the most stringent bound. For larger values of $\beta$, this restriction becomes less severe. Additionally, as discussed in relation to Eq.~\eqref{eq:Tst}, the upper limit $\beta < 1/2$ imposes a corresponding upper bound on $\Trh$, depicted by the red shaded region. As one can see from Eq.~\eqref{eq:Tstn2}, for a given $\alpha$, $\Tst$ reduces on increasing $\beta$. As a result, with increasing $\beta$, the contours shift to higher $\Trh$ to produce the right abundance. Based on the discussion in Sec.~\ref{sec:setup}, we show bounds from observations of the galactic and extra-galactic diffuse X-ray or gamma-ray background via the dashed vertical lines (regions to the right are allowed). As discussed in Sec.~\ref{sec:setup}, scalaron decaying entirely into a pair of $e^+e^-$ puts a lower bound on $\alpha\gtrsim 10^5\,\text{GeV}^{-2}$ (corresponding $\tau_\Phi\gtrsim 10^{24}$ s), while decay into photonic final states makes this lower bound even stringent $\alpha\gtrsim 4.3\times 10^7\,\text{GeV}^{-2}$ (corresponding $\tau_\Phi\gtrsim 10^{28}$ s). The bottom panel of Fig.~\ref{fig:paramspace} illustrates the viable parameter space for {\it Scenario-B}. In this case, we have the following set of free parameters that determine the resulting relic density allowed parameter space: $\{\lam,\,\mu,\,\gamma,\,\beta,\,\Trh\}$. Note that, the expression for the final DM yield remains exactly same as Eq.~\eqref{eq:yld-A}, excepting for the fact that now $\Tst$ is given by [cf. Eq.~\eqref{eq:TstB}],
\begin{align}  
\Tst\simeq 3\times 10^8\,\text{GeV}\,\sqrt{\frac{0.1}{\beta}}\,\left(\frac{10\,\text{GeV}^{-2}}{\lam}\right)^{1/4}\,.
\end{align}
 Once again, we see that the UV freeze-in characteristic is apparent, where a larger $\Trh$ requires lighter DM (a larger $\lam$ results in lighter DM as shown in Fig.~\ref{fig:lifetimeB}) to tame the overproduction. From the right panel of Fig.~\ref{fig:Tst}, one can see that $\Tst$ decreases with $\gamma$, with a maximum value of $\sim 10^9$ GeV for $\gamma=0.1\,\text{GeV}^{-2}$. We note, in both {\it Scenario-A} and {\it Scenario-B}, $\Tst<\Trh$ for our choice of parameters. For $\gamma\in\left[0.1-100\right]\,\text{GeV}^{-2}$, along with $\lam=-0.1\,\text{GeV}^2$ and $\mu=0.1$ GeV, the DM lifetime $4\times10^{17}\lesssim\tau_\Phi\lesssim 10^{23}$ s, which is slightly below the lower bound on lifetime from observation, typically for DM decaying into electron-positron pair.  
Thus, for these choice of the parameters, {\it Scenario-B} is in tension with bounds from diffuse X-/gamma-rays and as well as from CMB. This constraint, however, can be alleviated for $\mu\lesssim 10^{-2}$ GeV, which results in a lighter scalaron and eventually a longer DM lifetime $\tau_\Phi\gtrsim 10^{24}$ s.
\begin{figure}[htb!]
    \centering    
    \includegraphics[scale=0.52]{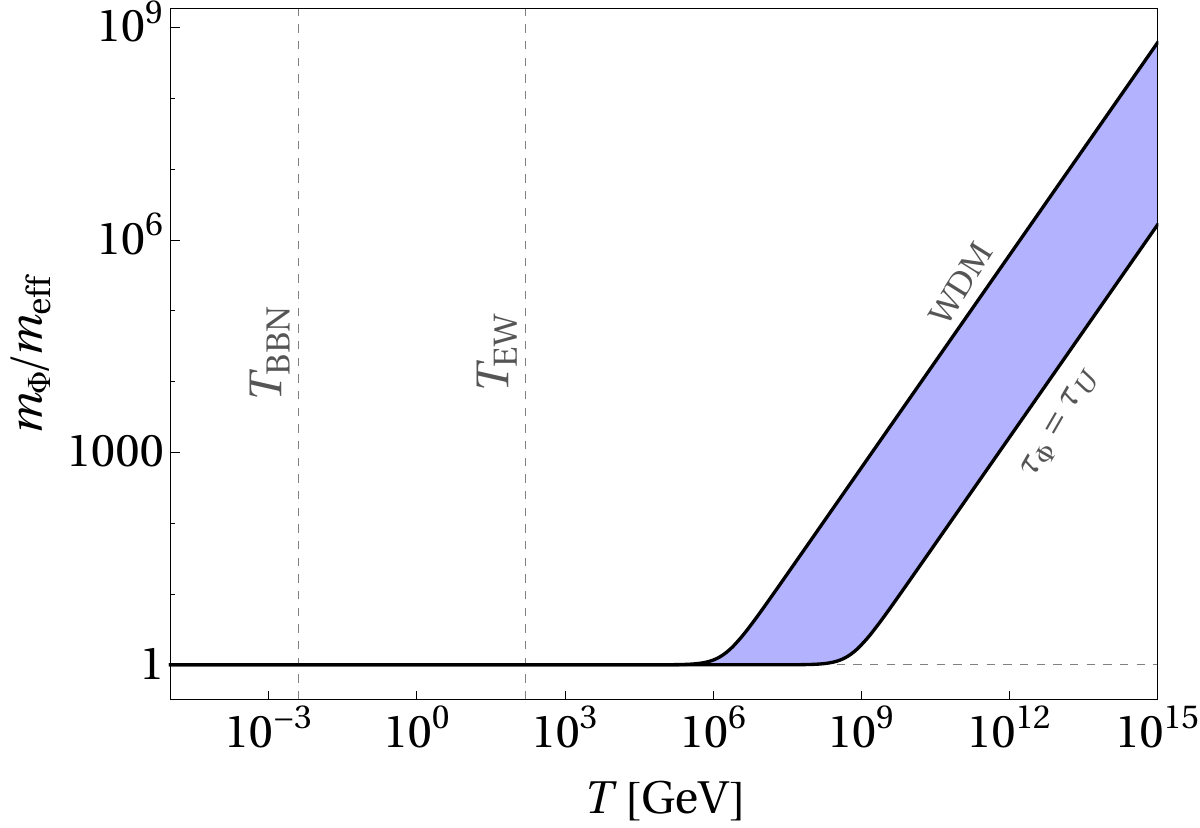}
    \caption{Comparison of scalaron bare mass $m_\Phi$ to the effective mass $m_{\rm eff}$ [cf.Eq.~\eqref{eq:meff}] as a function of temperature $T$ for $\beta=0.2$, considering {\it Scenario-A}. The lower and upper boundaries correspond to $\alpha$-values that result in cosmologically stable scalaron DM and bound from warm DM, respectively (same as in the upper panel of Fig.~\ref{fig:paramspace}). The blue shaded region is the viable range of scalaron mass, where it can be considered as a freeze-in DM candidate.}
    \label{fig:mcompare}
\end{figure}

Before closing this section, it is useful to comment on an important aspect of the scalaron mass. As shown in Refs.~\cite{Katsuragawa:2016yir,Katsuragawa:2017wge,Katsuragawa:2018wbe}, starting from the scalaron equation of motion one can construct an effective scalaron potential of the form, 
\begin{align}\label{eq:Veff}
    V_{\rm eff}(\Phi)
    = V(\Phi)
    - \frac{1}{4}\,e^{-4\sqrt{1/6}\,\kappa\Phi}\,T_\mu^{\ \mu}\,,
\end{align}
where \(V(\Phi)\) is defined in Eq.~\eqref{eq:vphi-fR}, and \(T_\mu^{\ \mu}\) denotes the trace of the energy--momentum tensor. The corresponding effective scalaron mass is then given by,  
\begin{align}\label{eq:meff}
m_{\rm eff}^2\equiv \left.\frac{\partial^2 V_{\rm eff}}{\partial\Phi^2}\right|_{\Phi=\Phi_{\rm min}}
= m_\Phi^2 + \frac{2\kappa^2}{3}\,e^{-4\sqrt{1/6}\,\kappa\Phi}\,T_\mu^{\ \mu}\,,
\end{align}
where \(m_\Phi^2 = \left.\frac{\partial^2 V}{\partial\Phi^2}\right|_{\Phi=\Phi_{\rm min}}\) is the bare scalaron mass. This explicitly shows that the visible-sector matter content (i.e., the SM energy density) modifies the scalaron potential, implying that the scalaron mass depends on the background matter distribution. The minimum of the effective potential satisfies  
\begin{equation}
\left.\frac{\partial V_{\rm eff}}{\partial \Phi}\right|_{\Phi_{\min}}=0\,,
\end{equation}
which leads to  
\begin{equation}
2 f(R_{\min}) - R_{\min} f_R(R_{\min})
+ \kappa^2 T^\mu{}_\mu = 0\,,
\label{eq:minimum_condition}
\end{equation}
where $R=R_{\rm min}$ denotes the the minimum. Using this condition, the scalaron mass can be written as,  
\begin{align}
m_{\rm eff}^2
= \frac{f_R - R f_{RR}}{3 f_R f_{RR}}\,,
\end{align}
where \(f_R(R)\equiv df/dR\) and \(f_{RR}(R)\equiv d^2 f/dR^2\). Let us now consider {\it Scenario-A}, for which one obtains  
\begin{align}\label{eq:meffR}
m_{\rm eff}^2
= \frac{1}{6\,\alpha\,(1+2\alpha R)}
\simeq
\begin{dcases}
\dfrac{t^2}{72\,\alpha^2\,\beta\,(2\beta-1)} 
& 2\alpha R \gg 1 \,, \\[10pt]
\dfrac{1}{6\alpha}
& 2\alpha R \ll 1\,,
\end{dcases}
\end{align}
where the first expression applies at early times, while the second corresponds to the late-time limit $R\to0$, and we have used Eq.~\eqref{eq:R}. Using Eq.~\eqref{eq:t}, one can determine the temperature $T_\times$ at which the condition $2\alpha\,R = 1$ is satisfied,
\begin{align}
T_\times = \sqrt{\frac{3}{2\pi}}\,\left[\frac{5}{\alpha\,\gs(T_\times)\,(1-2\beta)}\right]^{1/4}
\simeq 5.3\,\text{MeV}\,\left(\frac{2\times10^7}{\alpha}\right)\,,
\end{align}
which is around $T_{\rm BBN}\simeq 4$ MeV. It is evident that $T_\times$ decreases with increasing $\alpha$, indicating that the condition $2\alpha\,R = 1$ is fulfilled at progressively later cosmological times, while the dependence on $\beta$ is rather mild. Substituting Eq.~\eqref{eq:t} in Eq.~\eqref{eq:meffR}, we further find,
\begin{align}\label{eq:meff2}
& m_{\rm eff}^2\simeq m_\Phi^2\,\left[1-\frac{1-2\beta}{15\alpha}\,\sqrt{5\gs(T)\,\frac{T^4\,\alpha^3}{M_P^2}}\right]\,,
\end{align}
which shows, $m_{\rm eff}\to m_\Phi$ for $T\to0$. It is evident that the scalaron becomes lighter in the early Universe when the curvature is large. We further note that, according to Eq.~\eqref{eq:obsyield}, the observed DM abundance depends on $m_{\rm eff}(T_0)$, which coincides with the bare scalaron mass at the present epoch, as follows from Eq.~\eqref{eq:meff2}. This behavior is illustrated in Fig.~\ref{fig:mcompare}, where we compare the bare mass with the effective mass in {\it Scenario-A}. As shown, the effective mass is several orders of magnitude smaller than the bare mass at high temperatures, typically around the reheating temperature where the UV freeze-in occurs [cf. Fig.~\eqref{fig:paramspace}]. The effective mass approaches and eventually coincides with the bare mass well before the EW symmetry breaking temperature $T_{\rm EW}\simeq 160$ GeV. Thus, although the scalaron mass varies with the background curvature in the early Universe, this time dependence does not affect the determination of the present-day freeze-in abundance, which is fixed solely by the late-time (bare) scalaron mass.  
\section{Conclusion}
\label{sec:concl}
The $f(R)$ theory of gravity stands out as a compelling framework for describing cosmological evolution in a consistent way. Moreover, higher-order curvature terms such as $R^2,\,R_{\mu\nu}R^{\mu\nu}$ etc naturally emerge from quantum gravity corrections, providing strong theoretical motivation for studying the consequences of $f(R)$ framework. A particularly appealing feature of this theory is that it can be decomposed in terms of a pure gravitational sector, dictated by the Einstein's General Relativity and a scalar field, known as {\it scalaron}. Importantly, this scalar degree of freedom is not introduced by hand; rather, it arises inherently from the higher-curvature modifications of the gravitational action.

In the present work, we investigate the possibility of scalaron acting as a viable DM candidate. We employ the key merit of the $f(R)$ gravity that a scalar DM candidate appears naturally from the higher curvature effects of the underlying theory removing the need to introduce it by hand. Subsequently, we obtain the interaction of scalaron with the visible sector, where owing to its Planck-suppressed coupling strength with the visible sector, the freeze-in mechanism becomes a natural way to produce the DM non-thermally. The non-renormalizable nature of these interactions renders the DM production rate highly sensitive to the maximum temperature attained by the Universe, which we identify with the reheating temperature at the onset of radiation domination. By imposing observational constraints on the lifetime of scalaron DM, we find that satisfying the observed DM abundance requires a very high reheating temperature, $\Trh>10^{14}$ GeV. We translate this requirement into constraints on two well-known $f(R)$ gravity models, thereby placing bounds on the free parameters of these theories. In the power-law model, we restrict our analysis to the Starobinsky case with $n=2$, for which a physical minimum of the potential exists, giving rise to a well-defined bare mass term for the scalaron. In contrast, for $n\neq2$, the potential does not admit a physical minimum, thereby requiring one to instead consider the effective scalaron mass. As discussed, despite the effective mass being small in the early Universe, the DM abundance is governed by its present-day value. Therefore, the scalaron remains a viable DM candidate today, with its relic abundance produced via the freeze-in. It is important to emphasize that, unlike earlier studies such as Ref.~\cite{Cembranos:2008gj}, which examined constraints on an eV-scale scalaron produced through coherent oscillations, our analysis derives bounds on a MeV-scale scalaron generated via the freeze-in mechanism from the thermal bath. Moreover, this work provides insights into the maximal reheating temperature of the 
Universe, under the assumption of instantaneous reheating, consistent with avoiding scalaron DM overproduction. Finally, it is noteworthy that, unlike the chameleon mechanism where the scalaron mass is highly sensitive to the environment surrounding the scalaron field, the freeze-in 
mechanism works independent of the DM's effective mass.

A notable feature of this framework lies in the fact that (i) no additional beyond the Standard Model fields are introduced by hand, and (ii) the same modification to gravity also leads to a modified cosmological expansion history, particularly in the pre-BBN era. The only subtlety in the present framework is that the freeze-in mechanism operates successfully in the presence of an imperfect fluid, characterized by a non-vanishing energy-momentum trace, which nonetheless behaves like radiation in terms of its equation of state. Our results generally point toward a high-scale inflationary origin for MeV-scale scalarons produced via freeze-in during a radiation dominated epoch, independent of the underlying $f(R)$ model. A detailed investigation of scalaron production during post-inflationary reheating will be interesting and shall be pursued in a future work.  
\section*{Acknowledgment}
The authors acknowledge Alexander Belyaev and Alexander Pukhov for their help with the model implementation, Nicolás Bernal, Gaetano Lambiase and Sumanta Chakraborty for fruitful discussions. The authors thank Soumitra Sengupta for numerous insightful discussions, and Yuri Shtanov for providing very detailed feedback on the draft. 
\appendix
\section{Scalaron interactions}
\label{sec:int}
In the Einstein frame, we adopt unitary gauge for the Higgs doublet $\widetilde H\equiv\left(0~~~ h/\sqrt{2}\right)^T$. The covariant derivative is defined as
\begin{align}
\widetilde{D}_\mu = \partial_\mu-i\,g\,\left(\sigma^a/2\right)\,W_\mu^a-i\,g'\,\left(Y/2\right)\,B_\mu\,,
\end{align}
where $\sigma^a$ are the Pauli spin matrices $a=1,\,2,\,3$; $W(B)_{\mu}$ are the generators corresponding to $SU(2)_L[U_1(Y)]$ groups, $g^{(')}$ are corresponding coupling strengths and $Y=1$ is the Higgs hypercharge. From the action in Eq.~\eqref{action_ein_1} we can then write down the following interaction between the SM and the scalaron in the weak basis,
\begin{itemize}
\item {\underline{with gauge  kinetic term}}:
\begin{align}
&-\sqrt{\frac{2}{3}}\,\kappa\,\Phi\,\left|\widetilde{D}_\mu\,\widetilde{H}\right|^2=-\sqrt{\frac{2}{3}}\,\frac{\kappa}{2}\,\Phi\,\left(\partial_\mu h\,\partial^\mu h\right)-
\sqrt{\frac{2}{3}}\,\frac{\kappa}{8}\,\Phi\,h^2\,\Bigg[2\,g^2\,W_\mu^+\,W^{\mu-}
\nonumber\\&
-\left(g\,W_\mu^{(3)}-g'\,B_\mu\right)^2\Bigg]
-i\,\sqrt{\frac{2}{3}}\,\frac{\kappa}{2}\,\Phi\,\left(-g\,W_\mu^{(3)}+g'\,B_\mu\right)\,h\,\left(\partial^\mu h\right)
\,, 
\end{align}
where $W_\mu^\pm=\left(W^1_\mu\mp i\,W^2_\mu\right)/\sqrt{2}$.
\item {\underline{with fermion kinetic term}}:
\begin{align}
& i\,\kappa\,\sqrt{\f{3}{2}}\,\left(\Phi\,\bar{\widetilde{f}}\gamma^{\mu}\partial_{\mu} \widetilde{f}-\bar{\widetilde{f}}\gamma^{\mu}\,\partial_{\mu}\Phi\,\widetilde{f} \right)\,.    
\end{align}
\item {\underline{with Yukawa term}}:
\begin{align}
-2\,\kappa\,\sqrt{\frac{2}{3}}\,\Phi\,\widetilde{\mathcal{L}}_Y=-2\,\kappa\sqrt{\frac{2}{3}}\Phi\,\left(-y_f\,\widetilde{\bar f}_L\,\widetilde{f}_R\right)\,h+\text{H.c}\,,    
\end{align}
where $y_f$ is the corresponding Yukawa coupling strength.
\item {\underline{with Higgs potential}}: 
\begin{align}
& 2\,\kappa\,\sqrt{\frac{2}{3}}\,\Phi\,\widetilde{V}_H=2\,\kappa\,\Phi\,\sqrt{\frac{2}{3}}\,\left(-\mu_h^2\,h^2+\lambda_h\,h^4\right)\,,   \end{align}
where $\mu_h$ is the Higgs mass and $\lambda_h$ is the self coupling strength.
\end{itemize}
Before EWSB, we consider all the SM fields to be absolutely massless. Once the EW symmetry is broken (post-EWSB), the Higgs field obtains a non-zero vacuum expectation value (VEV) that can be parametrized as $ \left(0~~~(h+v)/\sqrt{2}\right)^T$, and the SM sector becomes massive. In the mass basis, 
\begin{align}
\begin{pmatrix}
A_\mu \\ Z_\mu
\end{pmatrix}
=\begin{pmatrix}
c_w & s_w\\
-s_w & c_w
\end{pmatrix}\,\begin{pmatrix}
B_\mu \\ W_\mu^3
\end{pmatrix}\,,
\end{align}
where $(c)s_w$ is the (co)sine of the Weinberg angle and $A(Z)_\mu$ is the physical photon (massive $Z$-boson). In Tab.~\ref{tab:vert} we have listed necessary scalaron-SM interaction vertices.
\begin{table}[htb!]
\centering
\begin{tabular}{c|c|c}\hline
Interactions & Before EWSB & After EWSB  \\
\hline\hline
$\Phi(p)-h(p_1)-h(p_2)$ & $p_1\cdot p_2$ & $2\,m_h^2-p_1\cdot p_2$ \\
\hline
$\Phi(p)-f(p_1)-f(p_2)$ & $3i\,\gamma^\mu\,\left(p-p_1\right)_\mu$ & $3i\,\gamma^\mu\,\left(p-p_1\right)_\mu-4\,m_f$ \\
\hline
$\Phi (p)-V(p_1)-V(p_2)$ & $-$ & $g^{\mu\nu}\,\epsilon_\mu(p_1)\cdot\epsilon_\nu(p_2)$ \\
\hline\hline
\end{tabular}
\caption{Lorentz structure of relevant scalaron-SM vertices obtained from Eq.~\eqref{action_ein_1}. Here $f$ and $V$ stands for SM fermions and gauge bosons, respectively.}
\label{tab:vert}
\end{table}
We implemented the model in LanHEP~\cite{Semenov:2008jy}, and obtained the relevant vertices and decay rates using CalcHEP~\cite{Pukhov:1999gg}.
\section{Scalaron decay rate}
\label{sec:DM-decay}
Pre-EWSB, considering all SM fields are massless, only possible two-body DM decay is into massless Higgs final state,
\begin{align}
& \Gamma_{\Phi\to hh}\simeq\frac{\kappa^2\,\mdm^3}{192\,\pi}\,.
\end{align}

Post-EWSB, where all the SM fields become massive, one finds,
\begin{align}
& \Gamma_{\Phi\to hh}=\frac{\kappa^2\,\mdm^3}{192\,\pi}\,\sqrt{1-4\,r_i^2}\,(1-6\,r_i^2)^2\,,
\nonumber\\&
\Gamma_{\Phi\to ff}=N_c\,\frac{28\,\kappa^2\,\mdm^3}{192\,\pi}\,r_i^2\,\left(1-4\,r_i^2\right)^{3/2}\,,
\nonumber\\&
\Gamma_{\Phi\to VV}=\delta\times\frac{\kappa^2\,\mdm^3}{192\,\pi}\,\sqrt{1-4\,r_i^2}\,\left(1-4\,r_i^2+12\,r_i^4\right)\,,
\end{align}
where $N_c=1(3)$ for leptonic (quark) final states and $r_i=m_i/\mdm$, $\delta=1(2)$ for $V=Z(W^\pm)$, and $m_i$ is the mass of the respective final state particle.

On top of the tree-level decay rates reported above, there are decays into gluonic and photonic final states that are generated at one-loop~\cite{Katsuragawa:2016yir,Shtanov:2025nue},
\begin{align}
\Gamma_{\Phi\to\gamma\gamma} &= \frac{\alpha_{\text{em}}^2 m_\varphi^3}{1024\,\pi^3 M_P^2} 
\left| b^{\text{eff}}(e) + a_W(\tau_W) + \sum_f N_c^{(f)}\,Q_f^2\,a_f(\tau_f) \right|^2, \\
\Gamma_{\Phi\to gg} &= \frac{\alpha_s^2 m_\varphi^3}{32\,\pi^3 M_P^2} 
\left| b^{\text{eff}}(g_s) + \sum_{f=\text{quarks}} a_f(\tau_f) \right|^2, 
\end{align}
where $Q_f$ is electromagnetic charge of the fermions, $\alpha_{\text{em}} = e^2/4\,\pi, \alpha_s = g_s^2/4\pi$, $\tau_{W/f}=4 m_{W/f}^2/m_\varphi^2$, with $e$ and
$g_s$ denoting the electromagnetic and QCD gauge couplings, respectively. Further, we have~\cite{Marciano:2011gm,Spira:2016ztx},
\begin{align}
a_W(\tau_W) &= -\left[ 2 + 3 \tau_W + 3 \tau_W (2 - \tau_W) f(\tau_W) \right], \\
a_f(\tau_f) &= \tau_f \left[ 1 + (1 - \tau_f)\,f(\tau_f) \right], \\
f(\tau) &= 
\begin{dcases}
\left[ \sin^{-1}\left( \frac{1}{\sqrt{\tau}} \right) \right]^2, & \tau > 1\,, \\
-\frac{1}{4} \left[ \ln \left( \frac{1 + \sqrt{1 - \tau}}{1 - \sqrt{1 - \tau}} \right) - i \pi \right]^2, & \tau \leq 1\,,
\end{dcases}
\end{align}
where, $b^{\rm eff}(e)=-11\,N_g/2,\,b^{\rm eff}(g_s)=-3\,N_g/2$, with $N_g=3$ being the number of generations.  
\section{Field equations in modified gravity}
\label{sec:field}
Considering the background metric to be
\begin{align}
& ds^2=-dt^2+a(t)^2\,d\vec x^2\,,    
\end{align}
the $00$ component of the field equation can be written as
\begin{align}\label{eq:00}
& \frac{f}{2}-3\,f'\,\left(\dot{\mathcal{H}}+\mathcal{H}^2\right)+18\,f''\,\left(4\mathcal{H}^2\,\dot{\mathcal{H}}+\mathcal{H}\,\ddot{\mathcal{H}}\right)=\kappa^2\,\rho\,,  
\end{align}
where the primes denote derivative with respect to $R$ and the dots are derivatives with respect to $t$. The Ricci scalar is given by
\begin{align}
R=6\,\left(\dot{\mathcal{H}}+2\,\mathcal{H}^2\right)\,,    
\end{align}
and,
\begin{align}
& \mathcal{H}=\frac{\dot a}{a}=\beta/t\,,
\dot{\mathcal{H}}=-\beta/t^2\,,
\ddot{\mathcal{H}}=2\beta/t^3\,,
\dddot{\mathcal{H}}=-6\beta/t^4\,,
\end{align}
where we have considered the scale factor $a(t)\propto t^\beta$, prior to the transition from $f(R)$-dominated to standard radiation dominated Universe. Using the time-scale factor relation we obtain,
\begin{align}\label{eq:R}
R = \frac{6\,\beta}{t^2}\,\left(2\beta-1\right)\,.    
\end{align}
Note that, for $\beta=0.5$, the Ricci scalar vanishes. This is expected since $\beta=0.5$ corresponds to the Universe entering a radiation-dominated phase, where the spacetime curvature sourced by matter vanishes due to the traceless nature of radiation. This is the expected behavior in GR and serves as a consistency check or matching condition for the transition from a modified to standard cosmology.

The field equation can be obtained by varying the action in Eq.~\eqref{action_fr_1},
\begin{align}\label{eq:field-eq}
& f'(R)\,R_{\mu\nu}-\frac{f(R)}{2}\,g_{\mu\nu}-\nabla_\mu\,\nabla_\nu\,f'(R)+g_{\mu\nu}\,\Box f'(R)=\kappa^2\,T_{\mu\nu}^{\rm matter}\,,    
\end{align}
where $T_{\mu\nu}^{\rm matter}$ is the energy-momentum tensor of the matter fields defined by
\begin{align}
& T_{\mu\nu}^{\rm matter} = -\frac{2}{\sqrt{-g}}\,\frac{\delta\mathcal{L}_{\rm matter}}{\delta g^{\mu\nu}}\,.     
\end{align}
The trace of Eq.~\eqref{eq:field-eq} is given by,
\begin{align}\label{eq:trace}
& 3\Box f'+f'\,R-2\,f=\kappa^2\,T\,,    
\end{align}
where
\begin{align}
\Box f'=-\ddot{f}'-3\,\mathcal{H}\,\dot{f}'\,,
\end{align}
and $T\equiv g^{\mu\nu}\,T_{\mu\nu}^{\rm matter}$. Einstein gravity, without the cosmological constant, corresponds to $f(R)=R$ and $f'(R)=1$, so that the term $\Box f'$  vanishes. In this case, we have $R = -\kappa^2 T$, and hence the Ricci scalar is directly determined by the trace $T$. 
\subsection{Scenario-A}
\label{sec:app-scenA}
The trace equation in Eq.~\eqref{eq:trace}, can be further reduced to
\begin{align}\label{eq:sc-A-trace}
& \alpha\,\Upsilon(\beta,n)\,R^n=\kappa^2\,T\,,    
\end{align}
using the modified scale-factor, where
\begin{align}
& \Upsilon(\beta,n)=\frac{\left(2\beta-n\right)}{\beta\,(2\beta-1)}\,\left[1-2\beta+n\,\left(2n+\beta-3\right)\right]\,.    
\end{align}
\begin{figure}[htb!]
    \centering        \includegraphics[scale=0.375]{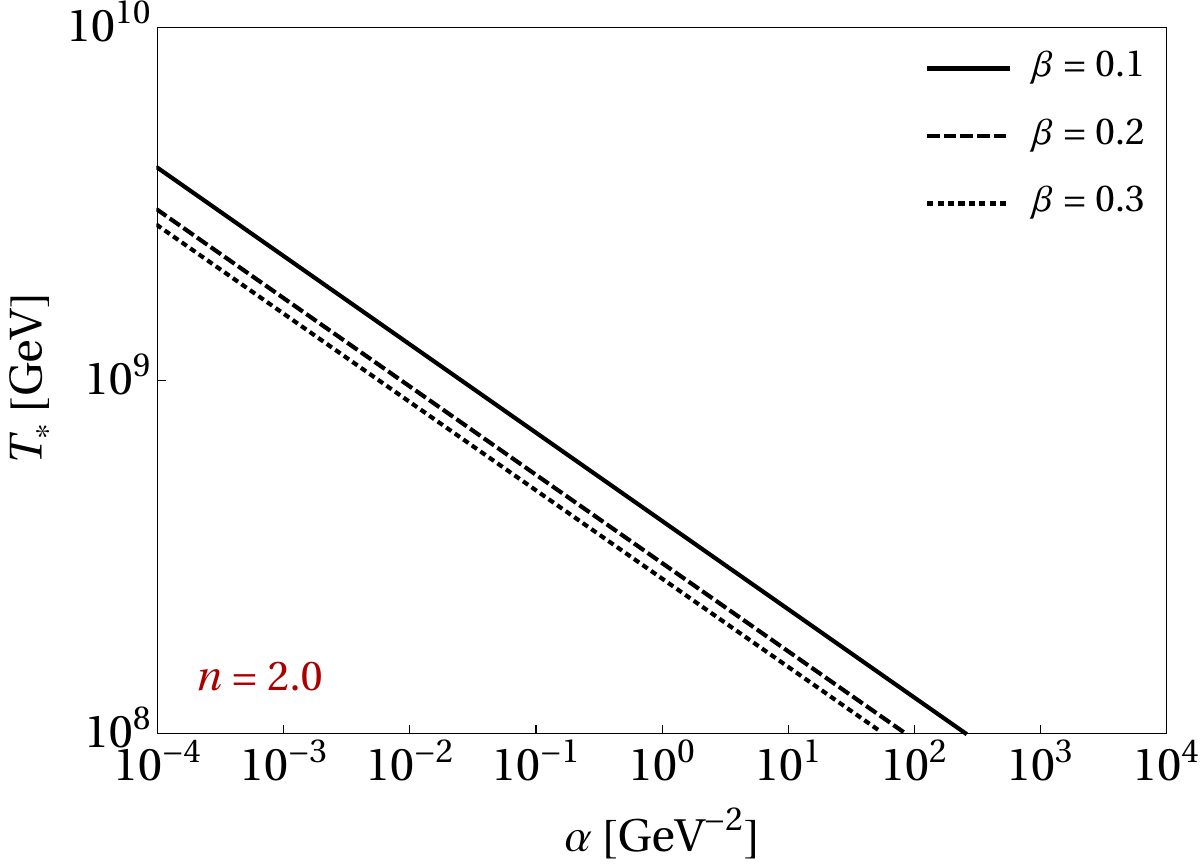}~    \includegraphics[scale=0.375]{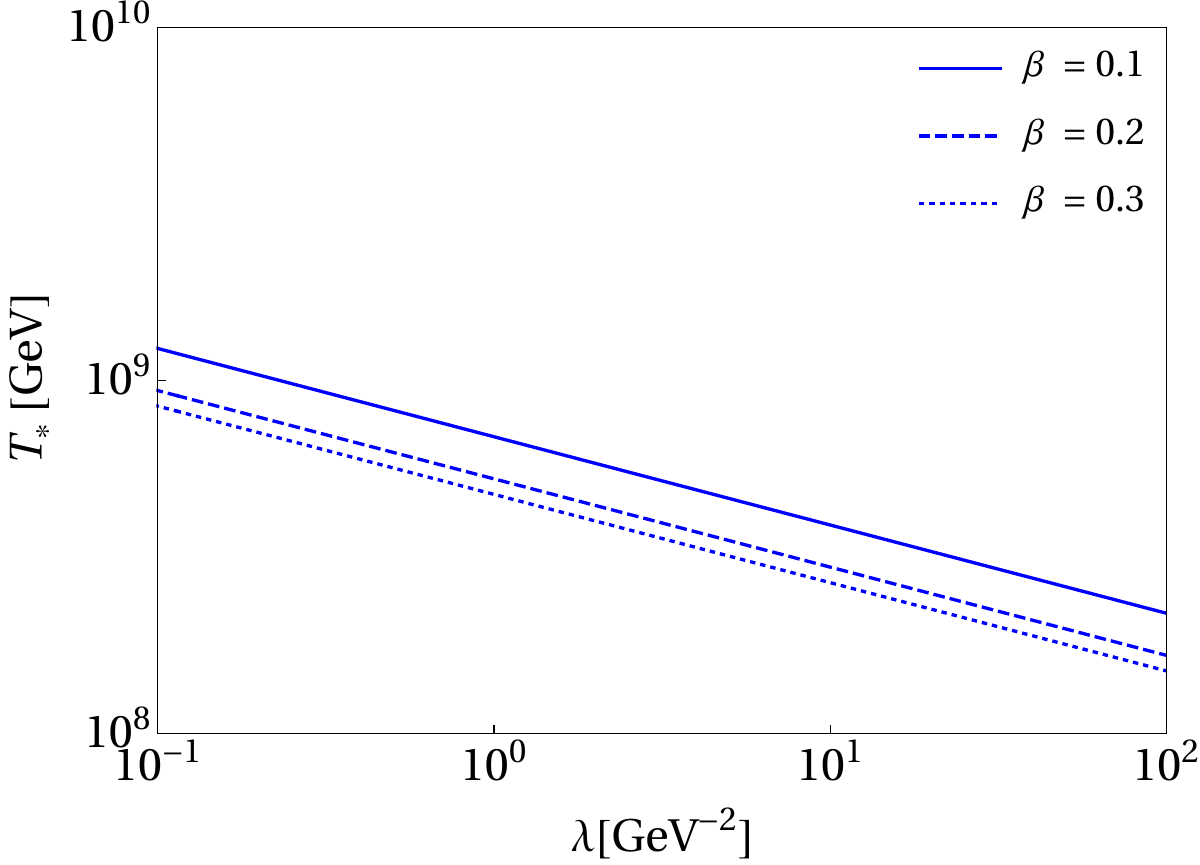}
    \caption{Left: {\it Scenario-A}: The transition temperature $\Tst$ as a function of $\alpha$ for different choices of $\beta$, following Eq.~\eqref{eq:Tst}, for $n=2$. Right: {\it Scenario-B}: Transition temperature $\Tst$ as a function of $\lam$ for different choices of $\beta$, following Eq.~\eqref{eq:TstB}.}
    \label{fig:Tst}
\end{figure}

Utilizing the fact $\rho_R=\left(\pi^2/30\right)\,\gs(T)\,T^4$, we obtain the following relation between the cosmic time $t$ and bath temperature $T$ from Eq.~\eqref{eq:00},
\begin{align}\label{eq:t}
&\alpha\,\Delta(\beta,n)\,R^n=\kappa^2\,\rho_R
\implies
t = \frac{z^{1/2}}{M_P}\,\left(\frac{T}{M_P}\right)^{-2/n}\,\left(\frac{30\,\Delta(\beta,n)}{\pi^2\,\gs(T)}\,\bar\alpha\right)^\frac{1}{2n}\,,    
\end{align}
with 
\begin{align}
& \Delta(\beta,n)=\frac{1}{2}\,\left(1-\left[\frac{n\,(\beta-3)+2n^2}{2\beta-1}\right]\right)\,,
\\&
\bar\alpha = \frac{\alpha}{M_P^{2\,(1-n)}}\,.
\end{align}
Here, $z=|6\beta\,(2\beta-1)|$, such that time is real for all $\beta$~\cite{Capozziello:2015ama}. The transition time $t_\star$ from modified to standard cosmology can be obtained by equating,
\begin{align}\label{eq:tst}
& \alpha\,\Delta(\beta,n)\,R^n(t_\star)=\mathcal H_{\rm st}(t_\star)^2\equiv\frac{1}{4t_\star^2}\implies t_\star=\frac{1}{M_P}\,\left(\frac{4}{3}\,\bar\alpha\,\Delta(\beta,n)\,z^n\right)^\frac{1}{2n-2}\,,   
\end{align}
where $\mathcal H_{\rm st}$ stands for Hubble during radiation domination, considering standard cosmological background. Corresponding transition temperature is given by
\begin{align}\label{eq:Tst}
T_\star=M_P\,\left(\frac{30}{\pi^2\,\gs(T_\star)}\right)^{1/4}\,\left[\left(\frac{4\,z}{3}\right)^n\,\bar\alpha\,\Delta(\beta,n)\right]^\frac{1}{4\,(1-n)}\,.   
\end{align}
For $n=2$, which is our choice of interest, the above equation simplifies to
\begin{align}
& \Tst=\frac{1}{2\sqrt{\pi}}\,\left(\frac{5}{\gs}\right)^{1/4}\,\sqrt{\frac{M_P}{\alpha^{1/2}}}\,\frac{1}{\left[\beta^2\,\left(1-2\beta\right)\right]^{1/4}}    
\end{align}
which is positive if $\beta<1/2$.
From Eq.~\eqref{eq:t} one can write,
\begin{align}
t=t_\star\,\left(\frac{T}{T_\star}\right)^{-2/n}\,.   
\end{align}
Therefore,
\begin{align}\label{eq:hub1}
\mathcal{H}=\frac{\dot a}{a}=\frac{\beta}{t_\star}\,\left(\frac{T}{\Tst}\right)^{2/n}\,.    
\end{align}
One can further write
\begin{align}\label{eq:hub2}
& \mathcal{H}(T)=\frac{\beta}{t_\star}\,\left(\frac{T}{\Tst}\right)^{2/n}\,\frac{\mathcal{H}_{\rm st}(T)}{\mathcal{H}_{\rm st}(\Tst)}\,\frac{\mathcal{H}_{\rm st}(\Tst)}{\mathcal{H}_{\rm st}(T)}=\frac{\beta}{\tst}\,\left(\frac{T}{\Tst}\right)^{2/n}\,\frac{\mathcal{H}_{\rm st}(T)}{1/\left(2t_\star\right)}\,\left(\frac{\Tst}{T}\right)^2
\nonumber\\&\implies
\mathcal{H}(T)=2\beta\,\left(\frac{T}{\Tst}\right)^q\,\mathcal{H}_{\rm st}(T)\,,~~~~~q=(2/n)-2\,,
\end{align}
where $\mathcal{H}_{\rm st}\propto T^2$ is the Hubble rate in standard cosmology. Clearly, there is an enhancement in the standard Hubble rate, that can be parametrized as
\begin{align}
\mathcal{E}(T)=2\,\beta\,\left(T/\Tst\right)^q\,. 
\end{align}
\subsection{Scenario-B}
\label{sec:app-scenB}
Keeping terms proportional to $R^2$ only, the trace equation in this scenario reads,
\begin{align}
& \frac{6\,\lam\,(1-\beta)}{\beta\,(1-2\,\beta)}\,R^2=\kappa^2\,T    
\end{align}
while the $00$ component of the field equation is given by,
\begin{align}
& \frac{3\,\lambda}{2\,(1-2\,\beta)}\,R^2=\kappa^2\,\rho^m\,,
\end{align}
where we have ignored terms that have logarithmic $R$-dependence, with respect to $R^2$. As before, we find time-temperature relation, 
\begin{align}
& t = \sqrt{\frac{z\,\lam^{1/2}}{M_P}}\,\left[\frac{90}{\pi^2\,\gs\,(2-4\,\beta)}\right]^{1/4}\,\left(\frac{T}{M_P}\right)^{-1}\,,     
\end{align}
while the transition time reads,
\begin{align}
& \tst = 2\,z\,\sqrt{\frac{\lambda}{2-4\beta}}\,,
\end{align}
with corresponding transition temperature,
\begin{align}\label{eq:TstB}
\Tst = \frac{M_P}{2}\,\sqrt{\frac{2-4\beta}{z}}\,\frac{1}{\sqrt{\lam^{1/2}\,M_P}}\,\left[\frac{90}{\pi^2\,\gs\,(2-4\beta)}\right]^{1/4}\,.  
\end{align}
Once again, we note that $\Tst>0$ for $\beta<1/2$. We therefore have,
\begin{align}
& t = \tst\,\left(\frac{\Tst}{T}\right)\,,~~~~\mathcal{H}=\frac{\beta}{\tst}\,\frac{T}{\Tst}\,.  
\end{align}
Therefore, the modified Hubble rate can be expressed as
\begin{align}\label{eq:hub-B}
&\mathcal{H}(T)=2\beta\,\mathcal{H}_{\rm st}(T)\,\left(\frac{\Tst}{T}\right)\implies\mathcal{E}(T)=2\beta\,\left(\frac{\Tst}{T}\right)\,.   
\end{align}
\bibliography{Bibliography}
\bibliographystyle{JHEP}
\end{document}